\documentclass[preprint2]{aastex}
\usepackage{emulateapj5}
\usepackage{onecolfloat5}
\input{pb.sty}
\shortauthors{Bhat et al.}
\shorttitle{Multifrequency Observations of Pulse Broadening}
\def\DM{{\rm DM}}     
\def\RM{{\rm RM}}
\def\SM{{\rm SM}}

\newcommand{\delDM}{\mbox{${\rm \delta DM}$}}

\newcommand{\dne}{{\mbox {$\delta n_e$}}}
\newcommand{\linn}{{\mbox {$l_{100}$}}}

\newcommand{\taudold}{{\mbox {$\tau_{d,tc93}$}}}
\newcommand{\taudnew}{{\mbox {$\tau_{d,ne2001}$}}}
\newcommand{\taudmean}{{\mbox {$\langle \tau_{d}\rangle$} }}
\newcommand{\cnsq} {{\mbox {$C_n^2$}}}
\newcommand{\Wtau}{\mbox{$W_{\tau}$}}

\newcommand{\be}{\begin{eqnarray}}
\newcommand{\ee}{\end{eqnarray}}

\begin{document}
\twocolumn[
\medskip
\medskip
\medskip
\medskip
\medskip
\title{Multifrequency Observations of Radio Pulse Broadening and
Constraints on Interstellar Electron Density Microstructure}
\medskip
\medskip
\author{N. D. Ramesh Bhat}
\affil{Massachusetts Institute of Technology, Haystack Observatory,
Westford, MA~01886 \\
rbhat@haystack.mit.edu}
\medskip
\author{James M. Cordes}
\affil{Astronomy Department and NAIC, Cornell University, Ithaca,
NY~14853\\
cordes@astro.cornell.edu}
\medskip
\author{Fernando Camilo}
\affil{Columbia Astrophysics Laboratory, Columbia University, 550 West
120th Street, New York, NY~10027 \\
fernando@astro.columbia.edu}
\medskip
\author{David J. Nice}
\affil{Department of Physics, Princeton University, Box 708, Princeton,
NJ~08544\\
dnice@princeton.edu}
\medskip
\author{Duncan R. Lorimer}
\affil{University of Manchester, Jodrell Bank Observatory, Macclesfield,
Cheshire, SK11~9DL, UK\\
drl@jb.man.ac.uk}

\medskip
\medskip
\begin{abstract}
We have made observations of 98 low-Galactic-latitude pulsars to
measure pulse broadening caused by multipath propagation through
the interstellar medium.  Data were collected with the 305-m Arecibo
telescope at four radio frequencies between 430 and 2380\,MHz.  We used
a CLEAN-based algorithm to deconvolve interstellar pulse broadening
from the measured pulse shapes.  We employed two distinct pulse broadening
functions (PBFs):  PBF$_1$ is appropriate for a thin screen of scattering
material between the Earth and a pulsar, while PBF$_2$ is appropriate
for scattering material uniformly distributed along the line of sight
from the Earth to a pulsar.  We found that some observations were
better fit by PBF$_1$ and some by PBF$_2$.  Pulse broadening times
($\tau_d$) are derived from fits of PBFs to the data, and are 
compared with the predictions of a smoothed model of the 
Galactic electron distribution. 
Several lines of sight show excess broadening, which we model as 
clumps of high density scattering material.  A global analysis of all 
available data finds that the pulse broadening scales with frequency, 
$\nu$, as $\taud \propto\nu^{-\alpha}$ where $\alpha\sim 3.9\pm 0.2$.  
This is somewhat shallower than the value $\alpha=4.4$ expected from 
a Kolmogorov medium, but could arise if the spectrum of turbulence 
has an inner cutoff at $\sim $300--800\,km.  A few objects follow particularly 
shallow scaling laws (the mean scaling index $\meanalpha \sim 3.1 \pm 0.1$ 
and $ \sim 3.8 \pm 0.2$ respectively for the case of PBF$_1$ and PBF$_2$), which 
may arise from large scale refraction or from the truncation of scattering 
screens transverse to the Earth--pulsar line of sight. 

\end{abstract}
]

\medskip
\medskip
\medskip

\section{Introduction} \label{s:intro}

\subsection{Overview} \label{s:overview}

Pulsars make excellent probes of the interstellar medium (ISM).  Observed
pulse profiles are influenced by dispersion, scattering, and Faraday
rotation along the line-of-sight (LOS) from the Earth to the pulsar.
Measurements of pulsars in similar directions at different distances can
be used to disentangle LOS interstellar effects and to model the ionized
content of the ISM \citep{tc93,bhat2002,CL2002a,CL2002b}.

We have undertaken multifrequency pulse profile observations using the
305-m Arecibo telescope, concentrating in the swath of the Galactic
plane visible from Arecibo, at Galactic longitudes $30^\circ\le l\le
75^\circ$.  The Parkes Multibeam Survey \citep[e.g.][]{manchesteretal2001}
has discovered hundreds of pulsars at low Galactic latitudes, $|b| <
5^\circ$, of which dozens are visible from Arecibo.  Many other pulsars
in this region are known from other survey work \citep[e.g.][]{ht1975b}.
Because the sensitive Multibeam Survey employed a higher frequency,
1400\,MHz, than most other surveys, the ``multibeam pulsars'' tend to be
relatively distant and highly scattered, making them particularly useful
for ISM studies.

Our most fundamental measurements are the set of pulse shapes at different
radio frequencies, from which we estimate the 
pulse broadening time scales caused by scattering,
\taud, for the pulsars.  In addition to providing input data to Galactic
electron density models, these measurements can be used to form an
empirical relation connecting \taud\ with dispersion measure, which
can serve as a useful guide in designing large-scale pulsar surveys
and in understanding the observable population of pulsars in the Galaxy
\citep[e.g.][]{bhattacharya1992,CL2002b}.

The paper is organized as follows.  Terminology and basic assumptions
about the ISM are summarized in \S~\ref{s:terminology}.  Details of
observations and data reduction are described in \S~\ref{s:obs},
and our method for deconvolving pulse broadening in \S~\ref{s:clean}.
Our results are presented in \S~\ref{s:der}~and~\S~\ref{s:pb}, and in
later sections we discuss the implications of pulse-broadening times for
the Galactic electron density models (\S~\ref{s:nemodel}), as well as
the power spectrum of electron density irregularities (\S~\ref{s:impli}).

\subsection{Terminology and Scattering Model} \label{s:terminology}

Several quantities measured by radio observations of a pulsar are
integrals of ISM properties along the LOS from the Earth to the pulsar.
The dispersion measure, $\DM \equiv \int _0 ^D ds\, \nele (s) $, is
the integral of electron density, \nele, along the LOS to the pulsar at
distance $D$.   Using a Galactic model for electron density,
\DM\ is often used to estimate pulsar distances.
We use cataloged values of $\DM$ in the analysis below to estimate distances.
The rotation measure, $\RM \equiv \int _0 ^D \nele(s) \,{\bf B}\cdot{\bf
ds}$, is the LOS integral of magnetic field, ${\bf B}$, weighted by
electron density.  Analysis of $\RM$ measurements from our data will be
reported in a future work.

Scattering of pulsar signals depends on fluctuations in the electron
density, $\dne$.  We assume that the spectral density of these
fluctuations follows a power-law model with cutoffs at ``inner''
and ``outer'' scales, \li\ and \lo, which are inversely related
to the corresponding wavenumbers, \ki\ and \ko, by $\li=2\pi/\ki$
and $\lo=2\pi/\ko$.  The spectral density is then given by
\citep[e.g.][]{rickett1977}:
\be
\Pne (\kappa) = \left\{
\begin{array}{ll}
\cn \kappa ^{-\beta},	& \quad\ko \le \kappa \le \ki \\
0,                      & \quad\mbox{elsewhere}. \\
\end{array}
\right.
\label{eq:spectrum}
\ee
The spectral coefficient \cn\ is expressed in
units of m$^{-20/3}$.  For Kolmogorov turbulence, the spectral slope is
$\beta =11/3$.

Pulse broadening is quantified by a time scale, \taud, characteristic
of a pulse broadening function (PBF) fit to a measured pulse shape.
The PBF is the response of the ISM to a delta function. The
exact form of the PBF and its scaling with frequency depend on the
spatial distribution of scattering material along the LOS and on its
wavenumber spectrum \citep{will1972, will1973, CR1998, LR1999, CL2001,
boldyrevgwinn2002}.  Therefore, determination of the PBF forms a useful
means for characterizing the underlying scattering geometry and wavenumber
spectrum for scattering irregularities.  The PBFs used in this work are
described in detail in \S\ref{s:clean}.

Measured pulse scattering parameters can be related to the scattering
measure, $\SM \equiv \int _0 ^D ds\, \cn (s) $, which is the LOS
integral of \cn.  For a Kolmogorov spectrum with a small inner
scale \citep[e.g.][]{rickett1990,CL1991,ARS1995}, the pulse broadening,
expressed as the mean arrival time of ray bundles \citep[see][]{CR1998},
is
\be
\taudmean &\approx& 1.1\, \Wtau\SM^{6/5}\nu^{-4.4}D,
\label{eq:taudmodel}
\ee
where $\nu$ is in GHz, $D$ is in kpc, SM is in \smu, \taud\ is in ms,
and \Wtau\ is a geometric factor that depends on the LOS distribution
of scattering material.

More generally, for a power-law wavenumber spectrum, the broadening time
scale follows a power law,
\be
\taud\ \propto \nu^{-\alpha}, 
\label{eq:nualpha}
\ee
where \citep[e.g.][]{cordesetal1986, romanietal1986},
\be
\alpha = \left\{
\begin{array}{ll}
{ 2 \beta \over (\beta - 2) } &  \quad \beta < 4 \\
{8 \over (6-\beta)}            &  \quad \beta > 4 \\
\end{array}
\right.
\label{eq:scaling1}
\label{eq:scaling2}
\ee
Thus, determination of $\alpha$ yields information about the wavenumber
spectrum. For a Kolmogorov spectrum, $\beta = 11/3$, implying
$\alpha=4.4$, and Eq.~\ref{eq:nualpha} reduces to \ref{eq:taudmodel}.
This result holds if the  inner scale of the spectrum is too small to
influence the measurements (Cordes \& Lazio 2002).  As we discuss later,
we infer that the inner scale likely does influence some of the scattering
measurements.

Finally, the decorrelation bandwidth, \nd, is related to \taud\ by $2 \pi
\taud\ \nd = C_1$, where the constant $C_1$, of order unity, depends on
the geometry of the scattering material and the form of the wavenumber
spectrum (Cordes \& Rickett 1998).

\section{Observations and Data Reduction} \label{s:obs}

The observations were made at the Arecibo Observatory.  New data for 81
pulsars were obtained in several observing sessions from 2001 May to 2002
November.  For the analysis in this paper we also use the data collected
by \citet{lorimeretal2002} for 17 pulsars, yielding a total
of 98 pulsars. We concentrated on pulsars
for which pulse broadening observations had not previously been made.
Prominent among these are 38 discovered in the Parkes Multibeam Survey
\citep[e.g.][]{manchesteretal2001}, 30 from the Hulse-Taylor survey,
including 17 with new timing solutions \citep{lorimeretal2002}, and 30
others \citep{tayloretal1993, hb03}

Data acquisition systems used for the observations are summarized in
Table~\ref{tab:obs}.  Signals were collected separately at four radio
frequencies, 430, 1175, 1475 and 2380 MHz.  The range of frequencies
was chosen to allow detection of pulse broadening over a wide variety of
pulsar scattering measures; specific frequencies were chosen according
to receiver availability and radio frequency interference environment.
The strong dependence of \taud\ on frequency implies that,  for most
objects, pulse broadening will be measurable at only a subset of the
four frequencies.   For pulsars with little scattering, pulse broadening
is detectable only at the lowest frequency, if at all.  By contrast,
for pulsars with heavy scattering,  broadening may be measurable at high
frequencies and may be so large as to render pulsations undetectable
at lower frequencies.  When pulsations are undetectable at 430 MHz,
the cause may also involve a combination of relatively small flux density
and large background temperature.

In the absence of any prior knowledge of flux density at the higher
frequencies, we adopted fixed integration times for all objects in a
first pass of observations.  Based on the initial results, one or more
re-observations were made during later sessions for those objects and
frequencies with low signal-to-noise ratios.

Observations at 430 MHz were made with the Penn State Pulsar Machine
(PSPM), an analog filterbank spectrometer providing 128 spectral channels
spanning an 8 MHz band in each of two circularly polarized signals.
Power measurements in each channel were synchronously averaged in real
time at the topocentric pulse period, yielding pulse profiles with time
resolution of approximately 1 milliperiod.
Dedispersion was done off line, reducing each observation to a single 
pulse profile.

Observations at 1175, 1475, and 2380 MHz were made with the Wideband
Arecibo Pulsar Processor (WAPP), a fast-dump digital correlator
\citep{dowdetal2000}.  Input signals to the WAPP were digitized into
three levels and output correlations were accumulated and written to disk
as 16-bit integers.  We recorded long time series of auto-correlation
functions (ACFs) and cross-correlation functions (CCFs) of the two
circularly-polarized polarization channels.  The ACFs were used in the
analysis of this paper.  A polarization analysis that utilizes both the
ACFs and CCFs will be reported in a subsequent paper.

In off-line analysis, the ACFs were van Vleck corrected
\citep[e.g.][]{hagen1973}, Fourier transformed, dedispersed, and
synchronously averaged to form average pulse profiles with, typically,
1 milliperiod resolution.  The software tools used to analyze the WAPP and
PSPM data are described by \citet{lor01b}.

Figure~\ref{fig:profs} shows the pulse profiles obtained from our
multiple frequency data.  Profiles for 5 pulsars are not shown due 
to poor data quality.  For the 37 multibeam pulsars shown, these 
represent the first observations at frequencies other than 1400\,MHz, 
and, in almost all cases, signal-to-noise ratios for the profiles 
are superior to those obtained in the original multibeam survey data.  
For nearly all 39 previously known pulsars in the sample shown here, 
the profiles are the best quality profiles obtained to date.

\section{Deconvolution Method}
\label{s:clean}

We used a CLEAN-based method \citep{bhatetal2003} for deconvolving
scattering-induced pulse-broadening from the measured pulse shapes.
This method does not rely on a priori knowledge of the pulse shape, and
it can recover details of the pulse shape on time scales smaller than
the width of the PBF.  A number of trial PBFs may be used, with varying
shapes and broadening times, corresponding to different LOS distributions
of scattering material.  The ``best fit'' PBF and broadening time are
determined by a set of figures of merit, defined in terms of positivity
and symmetry of the final CLEANed pulse, along with the mean and rms of
the residual off-pulse regions.  Details of the method and tests of its
accuracy are given in \citet{bhatetal2003}.

We used two trial PBFs.  The first, PBF$_1$, is appropriate for a thin
slab scattering screen of infinite transverse extent
within which density irregularities follow a
square-law structure function\footnote{The spatial structure function 
${\rm D_F(s)}$ 
of a quantity F(x) is defined as 
${\rm D_F (s) = \langle (F(x+s) - F(x))^2 \rangle}$ where s is the 
spatial separation (lag value).} \citep{LR2000}.
The PBF is given by a one-sided exponential \citep{will1972,will1973},
\be
{\rm PBF}_1(t) = \taud^{-1}\exp(-t/\taud) U(t), 
\label{eq:sharp}
\ee
where $U(t)$ is the unit step function, $U(<0) = 0, U(\ge 0) = 1$.
This function has been commonly used in previous pulsar scattering work.

The second broadening function, PBF$_2$, corresponds to a uniformly
distributed medium with a square-law structure function.  This PBF has
a finite rise time and slower decay,
\be
{\rm PBF}_2(t)=( \pi^5 \taud^3 / 4 t^5 )^{1/2} \exp (- \pi^2 \taud / 4t) U(t).
\label{eq:round}
\ee
This PBF is a generic proxy for more realistic distributions of scattering
material.

Additional PBFs, not used in our analysis, include those for media with
Kolmogorov wavenumber spectra, which can yield non-square-law structure
functions \citep[e.g.][]{LR1999}, and scattering screens that are 
truncated in directions transverse to the LOS, 
as may be the case for filamentary or sheet-like 
structures, which have PBFs that correspondingly 
are truncated at large time scales 
\citep{CL2001}.

Note that the pulse broadening time, \taud, has different meanings for
PBF$_1$ and PBF$_2$.  For PBF$_1$, \taud\ is both the $e^{-1}$ point
of the distribution and the expectation value of $t$.  For PBF$_2$,
\taud\ is close to the maximum of the distribution, which is at
$(\pi^2/10)\taud=0.99\tau_d$, while the expectation value of $t$ is
$(\pi^2/2)\taud=4.93\tau_d$.

For some of the pulsars we obtained an acceptable fit to \taud\
using both PBF$_1$ and PBF$_2$, while in others only one of the PBFs
provided an acceptable fit.   Acceptable fits were those that yielded
deconvolved pulse shapes that were positive, semi-definite; we reject
cases which yielded unphysical pulse shapes (such as profiles with
negative going components).  In many cases the pulse broadening is not
large enough to be measured, in which case we quote upper limits on \taud\
(see Table~\ref{tab:meas}).

As noted earlier in this section, our method relies on a set of figures of 
merit for the determination of the best fit PBF for a given choice
of the PBF form (see \citet{bhatetal2003} for details). Among the
different parameters used for this determination, the parameter
$f_r$ is a measure of positivity, and can serve as a useful indicator 
of ``goodness'' of the CLEAN subtraction.  However we emphasize that the 
absolute value of this parameter may also depend on the degree of 
scattering, the noise in the data, shape of the intrinsic pulse, 
etc., and therefore a comparison of the results for different
data-sets will not be meaningful. Nonetheless it can still be 
used for a relative comparison of the results obtained using 
different PBFs for a given pulse profile. For successfully 
deconvolved pulses, we expect $ f_r \la 1$; larger values 
imply slightly overCLEANed pulses. Based on this approach, 
the PBF with a lower value of $f_r$ can be considered to 
be the better of the two PBFs.

\section{Derived Intrinsic Pulse Shapes} \label{s:der}

\begin{figure*}
\epsscale{1.8}
\plotone{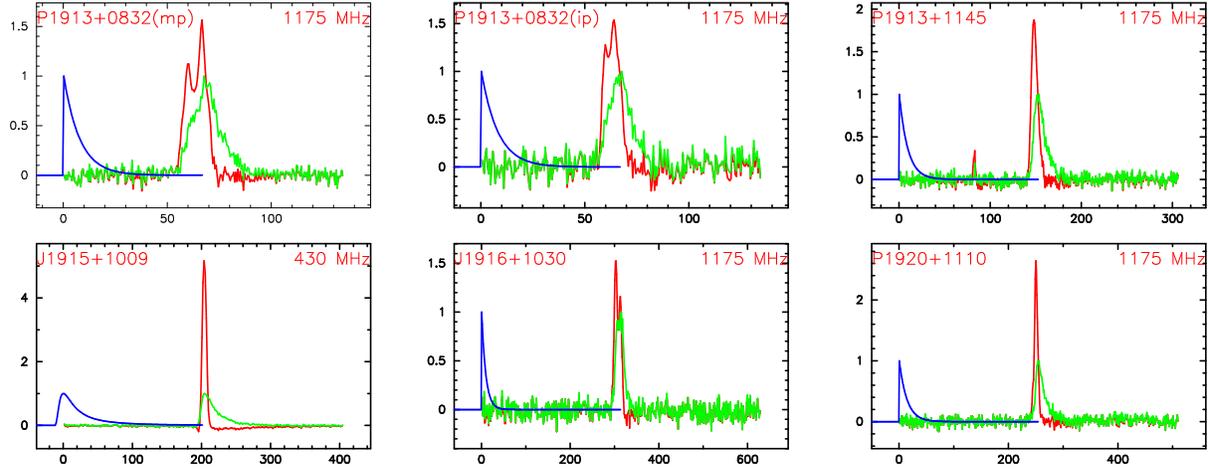}
\caption{Examples of the intrinsic pulse shapes  
(light solid [red] curves with highest peaks) 
and the best fit PBFs (solid [blue] curves rising 
from zero at left of each panel) obtained by 
application of the CLEAN method; the PBF is assumed 
to be a simple one-sided exponential (PBF$_1$, 
appropriate to a thin slab scattering geometry). The 
amplitudes of both the PBFs and the measured profiles 
(heavy solid [green] curves) are normalized to unity, 
and the areas under the intrinsic and measured pulse 
profiles are identical. }
\label{fig:eg-sharp}
\end{figure*}

\begin{figure*}
\epsscale{1.8}
\plotone{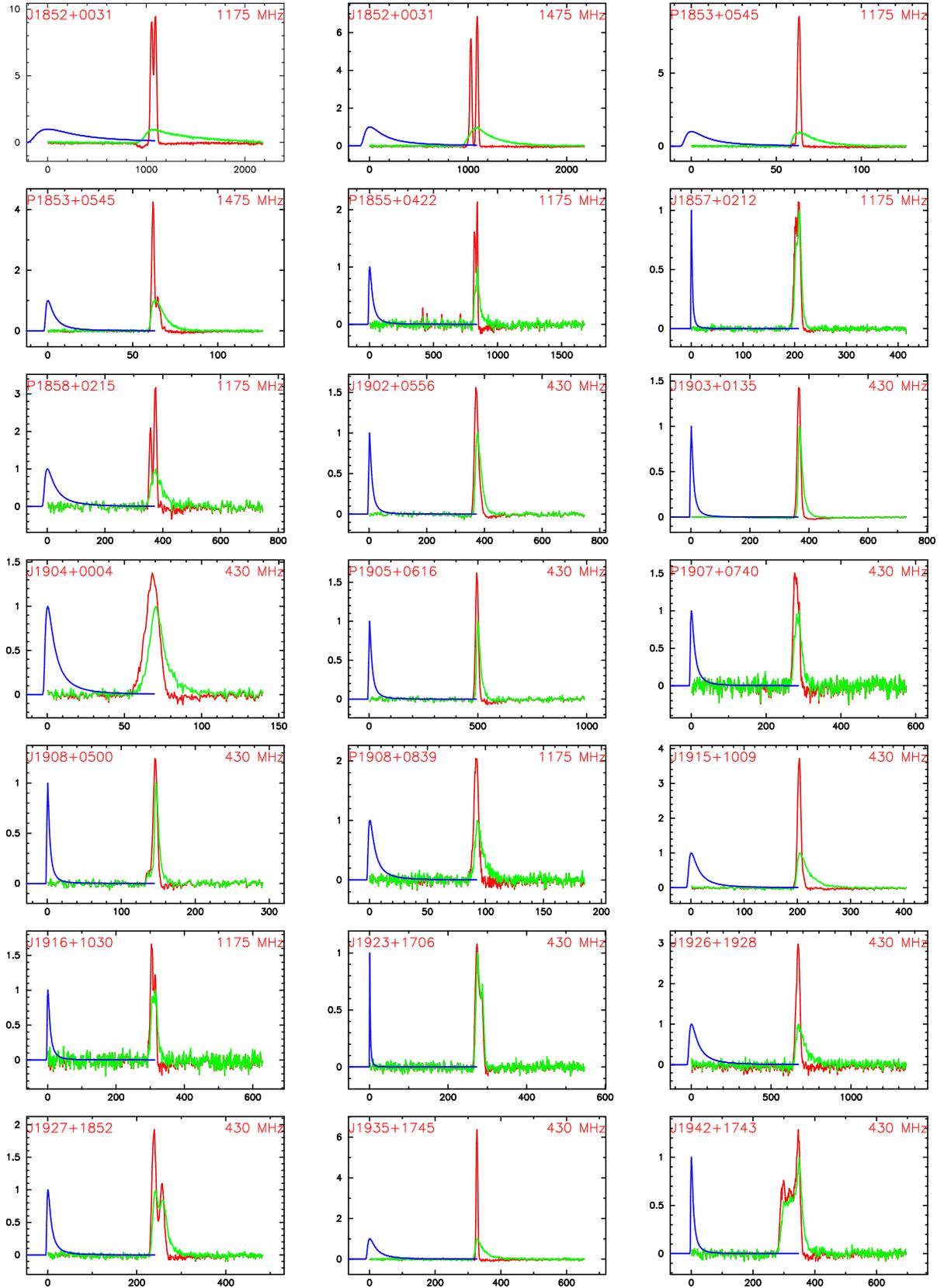}
\caption{Similar to Figure~\ref{fig:eg-sharp}, except that the pulse
broadening function employed by CLEAN has a more rounded shape (PBF$_2$,
due to a uniform scattering medium between the pulsar and the Earth). }
\label{fig:eg-round}
\end{figure*}

Figures~\ref{fig:eg-sharp} and \ref{fig:eg-round} show results from the
CLEAN-based deconvolution of our data.  In each panel, the best-fit
PBF is shown along with the observed pulse shape and the deconvolved
(intrinsic) pulse shape.  As is evident in the figures, the derived
pulse shapes are much narrower and significantly larger in amplitude
than the observed ones.  In several cases, the deconvolved pulse shapes
reveal significant structure which is not easily visible in the measured
pulse profiles.  For example, PSRs~J1913+0832 (Fig.~\ref{fig:eg-sharp})
and J1858+0215 (Fig.~\ref{fig:eg-round}) at 1175 MHz, have derived
pulse shapes that are distinct doubles, a property that is almost
entirely masked by broadening in the raw profiles.  In several other
cases (e.g. PSR~J1912+0828 at 1175 MHz, Fig.~\ref{fig:eg-sharp};
PSR~J1927+1852 at 430 MHz, Fig.~\ref{fig:eg-round}), the measured
pulse shapes show faint signatures of a double, which are confirmed
and reinforced by the deconvolution process. Data for PSR~J1906+0641
at 1175 MHz (Fig.~\ref{fig:eg-sharp}) and PSR~J1942+1743 at 430 MHz
(Fig.~\ref{fig:eg-round}) show that the technique yields details of
complex, multi-component pulse shapes.

While the deconvolution algorithm usually produces accurate pulsar
profiles, some cautions are in order.  As we already discussed,
for many objects successful deconvolution is possible using both
PBF forms.  Figures~\ref{fig:eg-sharp} and \ref{fig:eg-round} include
several examples of this kind.  In some cases, significantly different
intrinsic pulse shapes result from deconvolution with the two different
PBFs.  The data for PSR~J1852+0031 show an extreme example of this:
deconvolution with PBF$_1$ yields a pulse shape with three merged
components (Fig.~\ref{fig:eg-sharp}), while use of PBF$_2$ yields a
distinct double pulse shape (Fig.~\ref{fig:eg-round}).  Further examples
of substantial discrepancies include PSR~J1858+0215 at 1175 MHz, where use
of PBF$_2$ yields a double pulse, while use of PBF$_1$ yields a simpler,
single-peaked pulse profile, and PSR~J1853+0545 at 1175 MHz, where use of
PBF$_2$ yields a much narrower and more featureless profile than PBF$_1$.
However, such cases are exceptions rather than the rule.  There are many
examples where nearly identical intrinsic pulse shapes result with either
of the two PBFs. Data from PSRs~J1905+0616 and J1907+0740 at 430 MHz,
and J1908+0839 and J1916+1030 at 1175 MHz, belong to this category.

We emphasize that we have used only two extreme examples from the infinite
set of possible PBFs, and there may be LOSs for which other PBFs would
be more appropriate.  An exhaustive analysis using additional PBFs is
beyond the scope of this paper, though such an analysis may be valuable in
identifying important aspects of the ionized interstellar medium.  In the
remainder of the paper, we focus on measurements of pulse-broadening
times and their implications for models of Galactic free electron density.

\section{Pulse-broadening Times} \label{s:pb}

Our estimates of \taud\ are summarized in Table~\ref{tab:meas}.
The columns are: (1) pulsar name, (2) reference, (3) pulse period, (4)
\DM, (5) Galactic longitude, (6) Galactic latitude, (7) observation
frequency, (8) estimate of \taud\ using PBF$_1$, and (9) its figure 
of merit ($f_r$), (10) estimate of \taud\ using PBF$_2$, and (11) 
its figure of merit ($f_r$), (12) model estimate of pulse broadening 
using the TC93 model (\taudold), and (13) model estimate of pulse 
broadening using the NE2001 model (\taudnew).  
The definitions of model estimates are discussed below along with 
a comparison with measured values of \taud.  
We successfully measured \taud\ for 56 of the 98 target objects,
(of which 15 have measurements at more than one frequency), and 
obtained upper limits on \taud\ for 31 objects.

\subsection{Scaling of \taud\ with Frequency} \label{s:scaling}

For 15 pulsars, we have measured \taud\ at more than one frequency,
typically at 1175 and 1475 MHz, but in one case, PSR~J1853+0545,
also at 2380 MHz.  For 12 of these, estimates of \taud\ were possible
using both PBF$_1$ and PBF$_2$ (Table~\ref{tab:meas}),  and we derive
the estimates of the scaling index $\alpha$ in both cases (columns 4
and 6 in Table~\ref{tab:scaling}). We use \alphaone and \alphatwo to
denote the scaling indices for the two PBF cases, PBF$_1$ and PBF$_2$,
respectively, and the corresponding values for $\beta$ (obtained by use
of Eq.~\ref{eq:scaling2}) are denoted as \betaone and \betatwo.  For
PSR~J1853+0545, measurements of \taud\ are available for 3 frequencies,
and we estimate $\alpha$ for all three pairs of frequencies.

Despite the small sample of measurements, we find: (1) most cases show
significant departures from the traditional $\nu^{-4.4}$ scaling expected
for \taud, and (2) the inferred scaling index depends on the type of the
PBF adopted for the deconvolution. These have important implications
for the interpretations that will ensue in terms of the nature of the
wavenumber spectrum, as we discuss below.

\subsection{Scaling of \taud\ with DM } \label{s:tauvsdm}

An empirical relation connecting the pulse-broadening time and DM
serves as a useful guide in designing large-scale pulsar surveys.
An ideal pulsar survey will be scattering limited rather than
dispersion limited.
Most surveys to date have not, in fact, been scattering limited;
this is not because they have been poorly designed, but 
rather because they have been constrained by data throughput 
and computational limitations. In other words, scattering 
plays a significant role in determining the maximum distance 
to which a pulsar can be detected
and thus influences the observable population of pulsars. The relation
also provides some useful insights into the large-scale distribution of
free electrons (\nele) and the strength of their density fluctuations
(\delne) in the Galaxy.

\begin{figure*}
\plotone{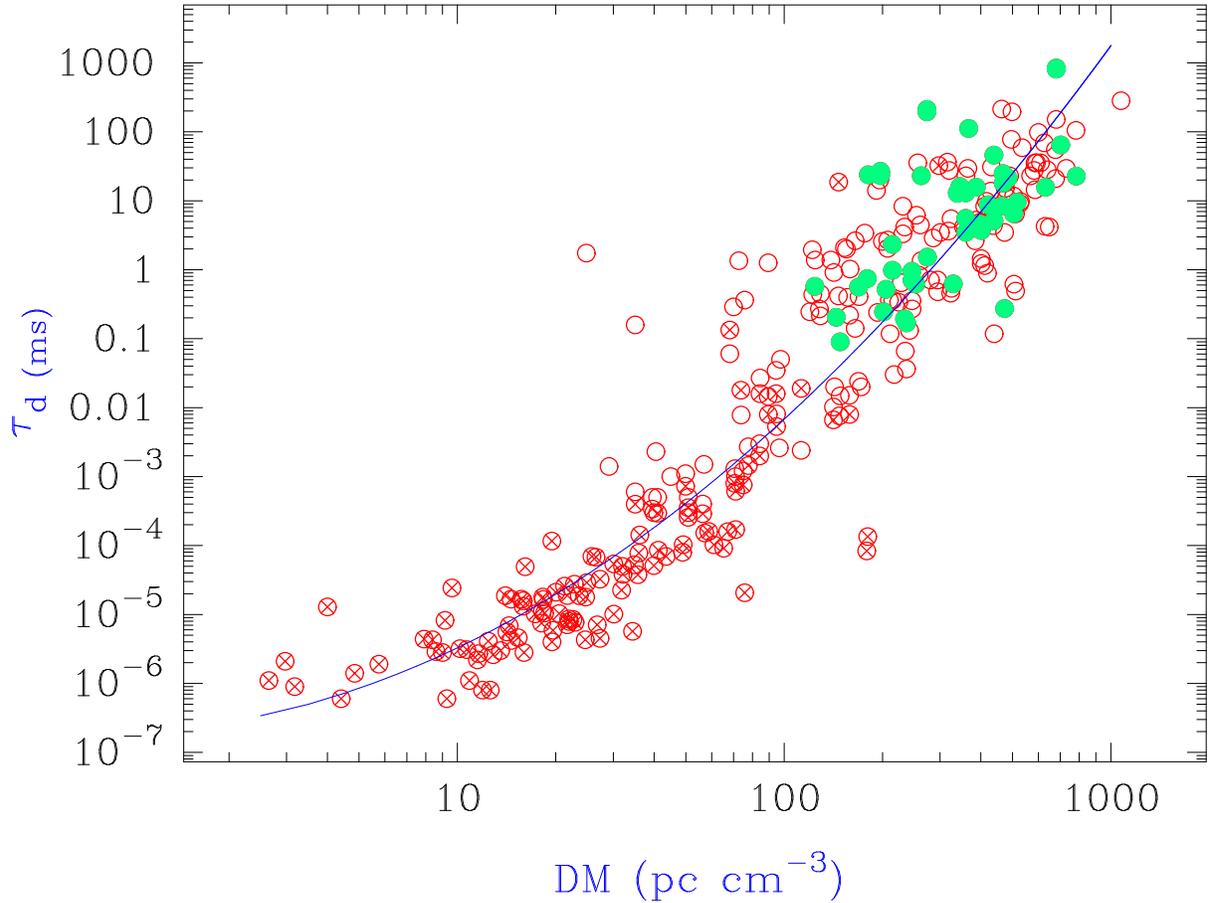}
\caption{Measurements of pulse-broadening times plotted against dispersion
measures.  The new measurements are shown as filled circles.  The open
circles with crosses (DM $\le$ 200 \dmu) are derived from the measurements
of decorrelation bandwidths, while the open circles are published \taud\
measurements.  The solid curve represents the best fit model for the
empirical relation between \taud\ and DM, the frequency-independent
coefficients for which are only slightly different from those obtained by
\citet{CL2002b} based on the published data alone (see \S~\ref{s:tauvsdm}
for details). }
\label{fig:tauvsdm}
\end{figure*}

Figure~\ref{fig:tauvsdm} shows a scatter plot of \taud\ and DM.
Most of the points at smaller DMs ($\la 100$ \dmu) are derived from
measurements of decorrelation bandwidth, \nd, which are converted
to scattering times by $\taud\ = C_1 / 2 \pi \nd$, assuming $C_1=1$.
Direct measurements of pulse-broadening times dominate at larger DMs
($\ga 100$ \dmu).  Evidently, there is  a strong correlation between
DM and \taud\ over the 10 orders of magnitude of variation in \taud.
The values of DM cover only 3 orders of magnitude, signifying a strong
dependence of pulse-broadening on DM.  There is also large scatter of
\taud\ about the trend,  roughly 2 to 3 orders of magnitude.  Some of
the scatter results from the fact that we have scaled all measurements
to a common frequency of 1 GHz using $\taud\ \propto \nu^{-4.4}$.
However, alternative scaling indices will yield an error of no more
than about 0.4 in $\log\taud$.  At lower DMs, some of this scatter
may be attributed to refractive scintillation effects which cause
fluctuations in the decorrelation bandwidth \citep[e.g.][]{bhat1999a}.
Also, some of the scatter may be due to the small numbers of ``scintles''
contributing to a measurement.  At larger DMs, the scatter is primarily
caused by strong spatial variations in \cnsq.

We fit the values of \taud\ and DM shown in Figure~\ref{fig:tauvsdm}
using a simple parabolic curve of the form \citep[e.g.][]{CL2002b}
\be
\log \taud \approx a + b (\log \DM) + c (\log \DM)^2 - \alpha \log \nu,
\label{eq:tauvsdm}
\ee
where $\nu$ is the frequency of observation in GHz, and \taud\ is
in ms.  Previous work has assumed a fixed scaling index, $\alpha =
4.4$, while fitting for the coefficients $a$, $b$ and $c$.  In the
light of our results discussed in \S~\ref{s:scaling}, and also other
recent work \citep[e.g.][]{lohmer2001} that suggest a departure from
the traditional $\taud\ \propto \nu ^{-4.4}$ behavior, we treat the
scaling index $\alpha$ as an additional parameter in determining
the best fit curve.  Note that most published measurements of \taud\
were determined by assuming a PBF of the form PBF$_1$ (and assuming
the conventional frequency-extrapolation approach). Hence we use our
values of \taud\ determined by using the same form of PBF (column 8 of
Table~\ref{tab:meas}) in order to ensure uniformity of the data used for
the fit.  Furthermore, to allow an unbiased fit for $\alpha$, we use
measurements in their {\it unscaled} form, i.e., direct estimates of
\taud\ and \nd\ at the observing frequencies\footnote{Note that many
published data, such as those in \citet{tayloretal1995}, are already
pre-scaled to a common frequency of 1 GHz.}. The data used for our fit, many
from prior compilations, include 148 estimates of \nd\ and 223 estimates
of \taud\ (of which 64 are our own measurements), thus 371 measurements
in total. Note that the upper limits are excluded from the fit, as none 
of them seem to impose any constraints to the fit.
For a subset of these objects, measurements are available at
multiple frequencies. The best-fit curve from our analysis is shown as
the solid line in Figure~\ref{fig:tauvsdm}.  Our re-derived coefficients,
$a=-6.46$, $b=0.154$ and $c=1.07$, are only slightly different from those
of \citet{CL2002b}, $a=-6.59$, $b=0.129$ and $c=1.02$. Interestingly,
the global scaling index derived from our best fit is $\alpha = 3.86\pm0.16$,
which is significantly less than the canonical value of 4.4  appropriate
for a Kolmogorov medium with negligible inner scale.  

There are several plausible explanations for departure from the
$\nu^{-4.4}$ scaling behavior for \taud, such as (i) the presence
of a finite wavenumber cutoff associated with an inner scale, (ii)
a non-Kolmogorov form for the density spectrum, and (iii) truncation
of the scattering medium transverse to the LOS, as addressed by
\citet{CL2001}.  Presently available observational data suggest that
option (i) may apply, so we investigate the effects of an inner scale on
the scaling laws for \taud.  Option (iii) may be relevant for specific
LOSs that contain filamentary or sheet-like structures that could mimic
truncated screens. In addition, there is yet another effect whereby a
weakening of the scaling index (as deduced from measurements of \nd\
and \taud) could result from refraction effects in the ISM. As argued
theoretically and demonstrated through observational data, refraction
from scales larger than those responsible for diffraction will bias
the diffraction bandwidth downward, corresponding to an upward bias on
pulse-broadening \citep[e.g.][]{cordesetal1986,gupta1994,bhat1999b}.
The refraction effects will be stronger at higher frequencies as one
approaches the transition regime between weak and strong scattering,
which will tend to weaken the frequency dependence from 4.4 to a lower
index. For pulsars at low DMs (say, DM $\la$ 100 \dmu), this transition is
expected near $\sim$1--3 GHz. Our sample contains many low-DM objects 
with measurements at $\sim$1--2 GHz where such an effect may be 
significant. 



\subsubsection{Effect of finite inner scale on $\alpha$}
\label{s:inner}

The presence of a finite inner scale can potentially modify the
frequency scaling index as estimated from measurements of \nd\ and
\taud.  \citet{CL2002b} show that these effects become apparent above a
``crossover point'' that is a function of distance (or DM) as well as
the observing frequency $\nu$. The crossover point can be defined for
commonly used observables such as \thetad (angular broadening), \nd\
and \taud.  In order to examine our data for any such signatures of an
inner scale, we define a ``test quantity'' in terms of \taud, $\nu$,
and distance ($D$) that is directly related to the inner scale, expressed 
in units of 100\,km, $\linn= l_i/(100\,{\rm km})$.  The crossover point
\taudcross\ is related to the inner scale by 
(see Eq.~A20 of \citet{CL2002b}),
\be 
\taudcross\ \approx {\rm 5.46 ~ms~} D (\nu _{\rm GHz} \linn )^{-2}, 
\label{eq:taudcross}
\ee 
where $D$ and $\nu$ are in kpc and GHz, respectively. Thus, a useful 
test quantity for identifying a break point in the frequency scaling 
is $ \taudcross\ \nu^2 / D $. In the analysis that follows, we use a
simple linear relation to convert DM measurements to distances,
$D=\mbox{DM}$/(1000\avne ), where \avne=0.03~\neu is the mean electron density,
and DM is in units of \dmu. 
We emphasize that we adopt such a simplistic approach as a 
preliminary step, and will defer to another paper a more 
detailed and complete analysis using proper electron density 
models and the independent pulsar distance estimates.

We split the data-set into two parts, below and above a chosen break point
value for this test quantity, and for each case we refit the parabolic
curve in Eq.~\ref{eq:tauvsdm} for the best fit $\alpha$ while keeping the
coefficients $a$, $b$, $c$ fixed at their global fit values. We do this
exercise for several break point values in the range 0.03--3.3, determining
the difference in best fit $\alpha$ values for the two samples in each
case ($\dalpha = \alphamin - \alphaplus$, where \alphamin and \alphaplus
denote the values of $\alpha$ for the samples that are below and above
the break point).  If an inner scale effect is truly relevant, we will
expect a significant difference in $\alpha$ for the two samples (with a
larger value for the sample below the break point, i.e., $ \alphamin >
\alphaplus $).

\begin{figure*}
\epsscale{0.9}
\plotone{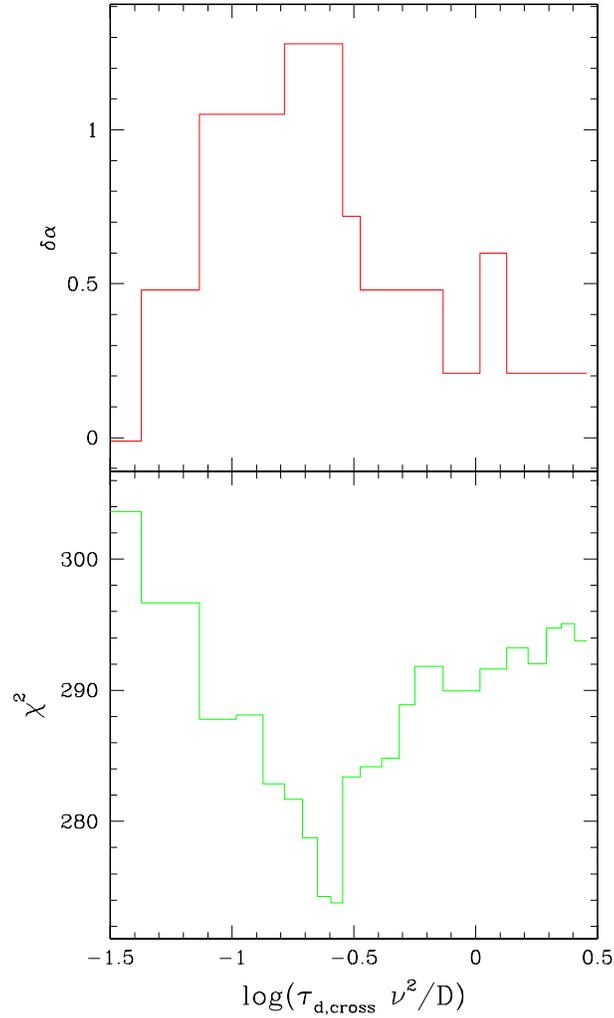}
\caption{Analysis of the frequency dependence of pulse broadening
that takes into account an inner scale for the wavenumber spectrum of
electron-density irregularities.  ({\em Top panel\/}) Plot of $\dalpha$,
the difference in exponent in the relation $\taud\propto\nu^{-\alpha}$
above and below a break point defined by the composite quantity
$\testqty$.  We calculate the best fit values of $\alpha$ for data points
above and below the break point and calculate $\dalpha$ as a function of
$\testqty$.  The units of $\testqty$ are (ms GHz$^2$ kpc$^{-1}$).
({\em Bottom panel\/})  $\chisq$ for the fit as a function of $\testqty$,
defined here as the sum of the squares of data$-$model (see text).}
\label{fig:inner}
\end{figure*}

Figure~\ref{fig:inner} shows a plot of \dalpha\ vs. the test quantity
\testqty, along with a corresponding plot of the best fit $\chi^2$
($\chisq = \chisqone + \chisqtwo$, where \chisqone and \chisqtwo
denote the corresponding values of $\chi^2_i$ for the two data-sets).
The maximum in \dalpha\ roughly coincides with the minimum in \chisq,
suggesting that the inner scale effect is real.
Our analysis shows a sharp minimum for \chisq\ at $\log(\testqty)\approx
-0.57$ (Fig.~\ref{fig:inner}).  Formally, the $\pm 1\sigma$ error on
the break point value of $\log(\testqty)$ is $\pm 0.05$.  However,
the valley in $\chi^2$ is much broader than implied by this error.
We take a more realistic range to be $-1$ to $-0.3$ in the log,
corresponding to an inner scale $l_i \approx 100\,{\rm km}\, \left
(5.46 \,D / \taud\ \nu^2\right)^{1/2} \approx 300$ to 800 km. 

The broadness of $\chi^2$ is caused in part by our assumption of a
simple proportionality between distance and \DM\ and also by the likely
variation of inner scale between locations in the Galaxy.   Some theories
for density fluctuations in the ISM would associate the inner scale with
the proton gyroradius for thermal gas.  The gyroradius is $r_g \approx
1658\,{\rm km}\, T_4^{1/2} B_{\mu\rm G}^{-1}$ for a temperature $T =
10^4 T_4$K and a magnetic field strength $B$ expressed in micro gauss.
For ionized gas in the warm phase of the ISM, we expect the temperature
to vary by a factor of 2 to 4 and the field strength by at least a
similar factor.  Thus, we would expect the gyroradius to vary by at
least a factor of five, which is not inconsistent with the appearance
of $\chi^2$ in Figure~\ref{fig:inner}.  Given the expected variation of
the gyroradius in the ISM, it is perhaps surprising that we see any kind
of minimum in $\chi^2$ at all.

Several authors have investigated the effect of an inner scale,
and constraints are available from various kinds of observations.
For example, \citet{spanglergwinn1990} derived an inner scale
of $\sim$50--200 km from an analysis of interferometer visibility
measurements from VLBI observations, which are, interestingly, of the
order of our estimates derived from pulse-broadening data. Further,
as noted by \citet{moran1990}, observations of NGC~6334B (the object
with the largest known scattering disk) at centimeter wavelengths are
consistent with an inner scale larger than 35 km. Studies of long-term
flux density variations of pulsars at low radio frequencies, however,
indicate a much larger inner scale (e.g.~$\sim10^2$--$10^4$ km from the
work of \citet{gupta1993}). While some discrepancies prevail between the
estimates deduced from different observations, it appears that effects
due to an inner scale are well supported by a number of observations.

\section{Galactic Electron Density Models} \label{s:nemodel}

Our sample largely comprises high-DM, distant pulsars and hence provides
useful data for improving upon electron density models for the inner
parts of the Galaxy.  We compare our data with predictions from both
the TC93 \citep{tc93} and NE2001 \citep{CL2002a, CL2002b} models,
which yield values for \SM\ that may be used in Eq.~\ref{eq:taudmodel}.
The newer model, NE2001, has made use of only \DM\ values of some of the
multibeam pulsars; hence, our measurements of \taud\ allow an independent
test of the new model.

\begin{figure*}
\epsscale{1.8}
\plotone{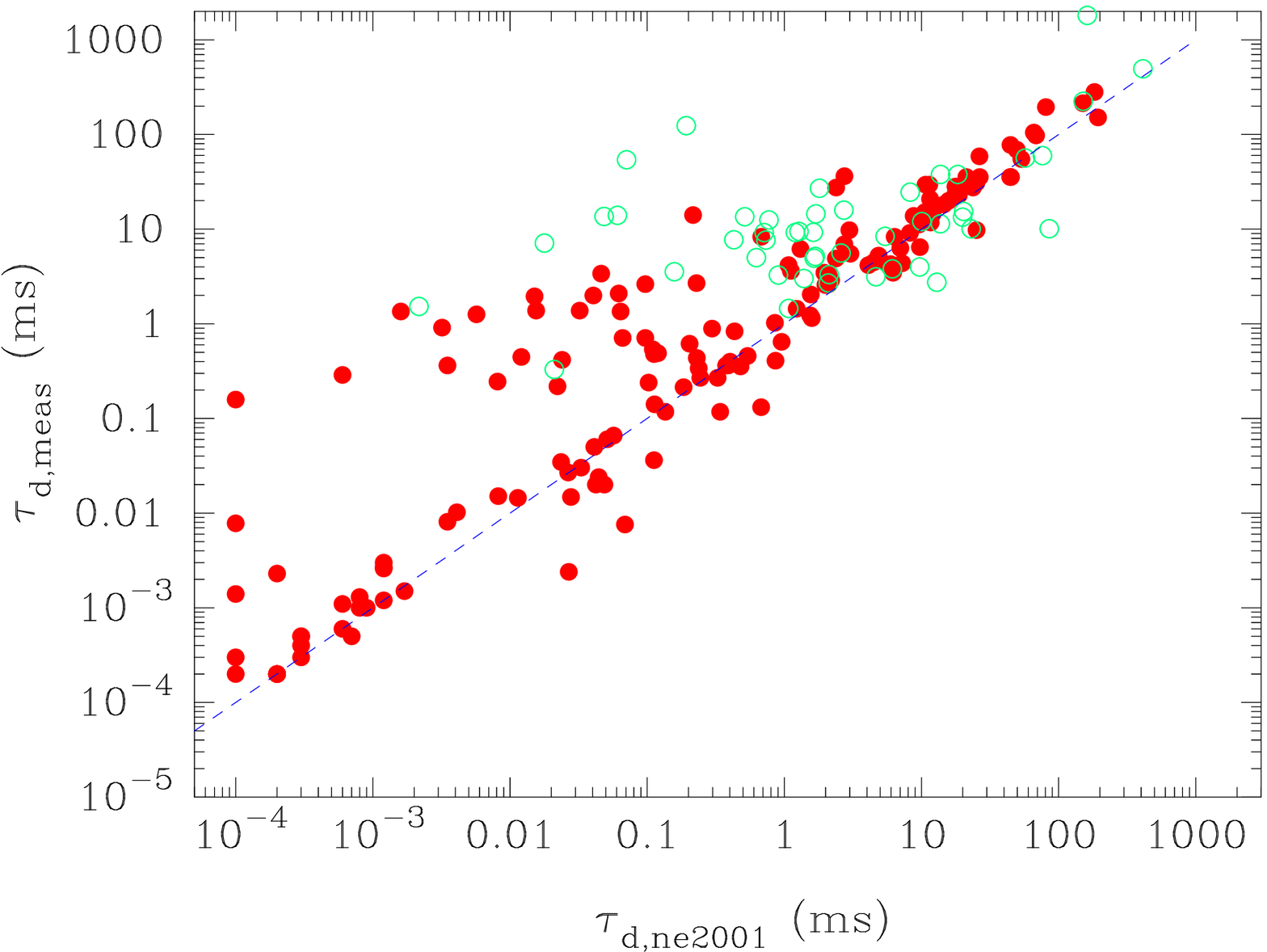}
\caption{Measurements of pulse-broadening times plotted against the
predictions from the new electron density model NE2001 \citep{CL2002a}.
The filled circles are the published measurements. The new measurements
from our observations are shown as open circles. All measurements
are scaled to a common frequency of 1 GHz using $\taud \propto
\nu^{-4.4}$. The dashed line is of unity slope. As evident from the
figure, a significant number of both the published and new measurements
are well above the dashed line, which implies that the model tends to
underestimate the degree of scattering toward many lines of sight. }
\label{fig:tau-ne2001} 
\end{figure*}

\begin{figure*}
\epsscale{1.8}
\plotone{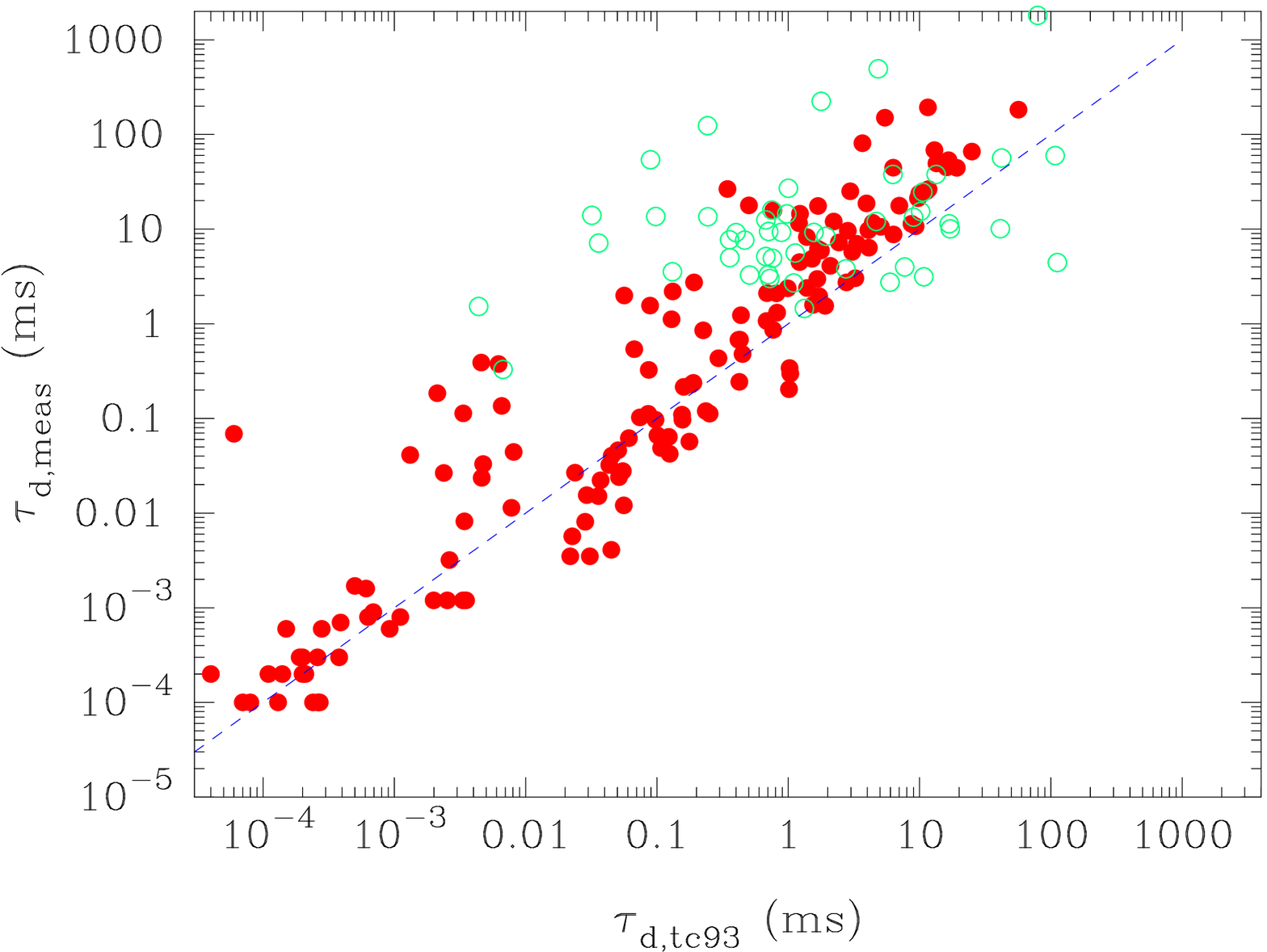}
\caption{Measurements of pulse-broadening times plotted against the
predictions from the electron density model TC93 \citep{tc93}.  The filled
circles are the published measurements, and the new measurements from
our observations are shown as open circles.  All measurements are scaled
to a common frequency of 1 GHz using $\taud \propto \nu^{-4.4}$. The
dashed line is of unity slope. As for Figure~\ref{fig:tau-ne2001}, the
model tends to underestimate the degree of scattering toward many lines
of sight. }
\label{fig:tau-tc93} 
\end{figure*}

Figures~\ref{fig:tau-ne2001} and \ref{fig:tau-tc93} show plots of the
measurements of pulse-broadening times against the predictions from
the new and old electron density models, respectively.  In order to
examine more general trends, we also plot all the published measurements
(see \citet{tayloretal1995} and references therein), after scaling to
a common frequency of 1 GHz using $\taud\ \propto \nu^{-4.4}$.  
A significant number of measurements show reasonable agreement 
with the model predictions, suggesting that the models depict 
fairly good representation of the large-scale picture in the 
Galaxy. However, significant discrepancies are evident in many 
cases, compared against the predictions from either of the two 
models.  For a majority of the measurements ($\sim$75\%) from
our own observations, the discrepancy is significantly lower 
with the predictions of NE2001 than with those of TC93. In 
some cases, the agreement with the model prediction shows 
improvements of the order a factor of two or better. 
Given that our measurements were not part of the inputs 
for the new model, this comparison makes an independent 
test of the new model. 

\subsection{Clumps of Excess Scattering} \label{s:clumps}

As discussed in \S~\ref{s:terminology}, 
the measured broadening time \taud\ is related to the total amount 
of scattering, usually quantified
as the scattering measure, SM (see Eq.~\ref{eq:taudmodel}). For a given
scattering geometry (indicated by the corresponding geometric factor
$W_\tau$ in the equation), we can invert this equation to derive
the scattering measure. We assume $W_\tau$=1 and estimate the {\em
effective\/} SM for a uniform medium. The estimated values of SM 
(\smobs, in the conventional units of \smu) are listed in 
Table~\ref{tab:sm} (column 7).

\begin{figure*}
\epsscale{1.8}
\plotone{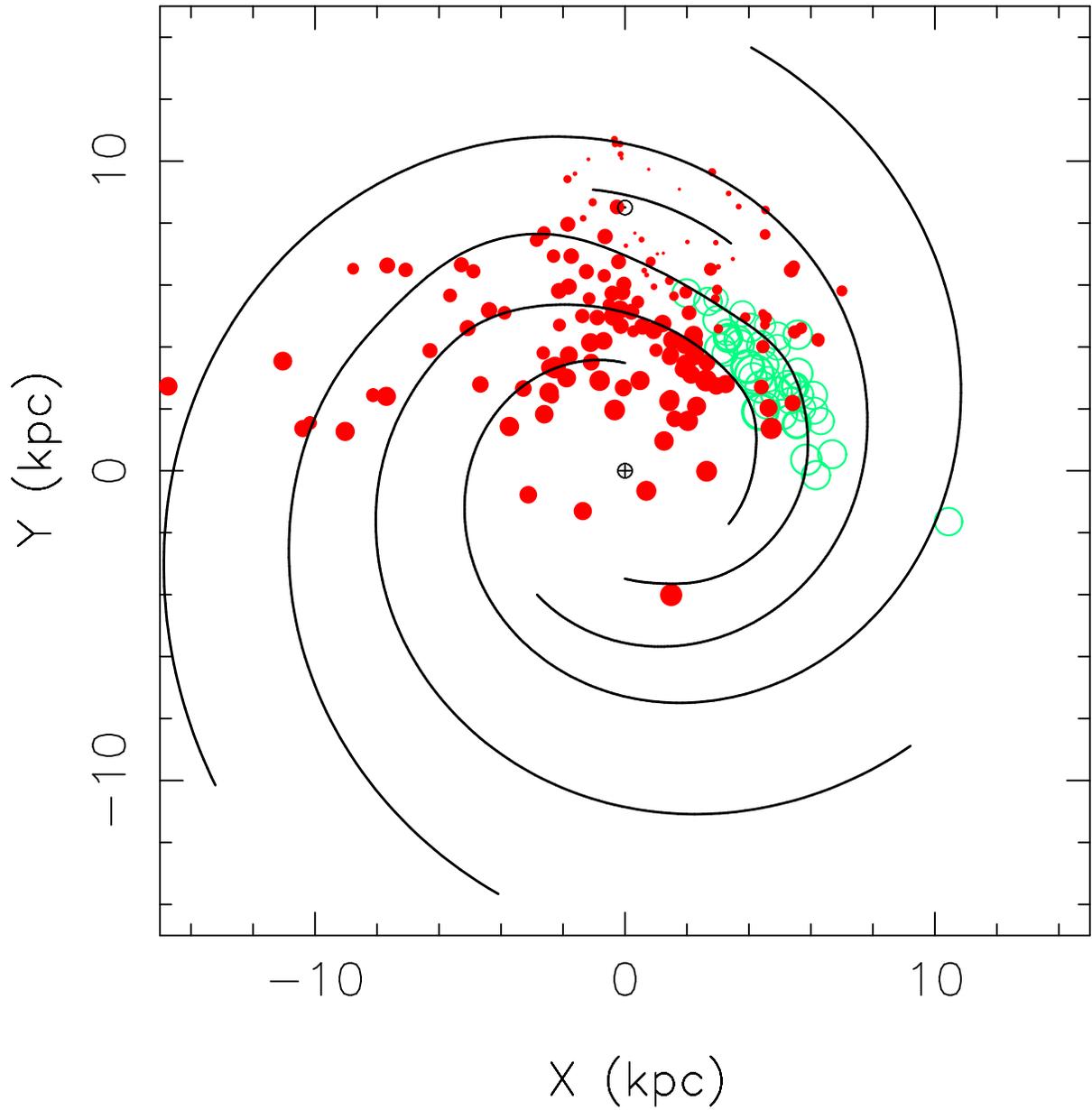}
\caption{Estimates of scattering measure (SM) derived from all
pulse-broadening data available. The size of the symbol is 
proportional to log(SM). Pulsar positions are projected onto
the Galactic plane; filled circles represent the published data,
and the new measurements from our observations are shown as open
circles. The spiral arm locations are adopted from the NE2001 model of
\citet{CL2002a,CL2002b}.}
\label{fig:xyplot}
\end{figure*}

\begin{figure*}
\epsscale{1.8}
\plotone{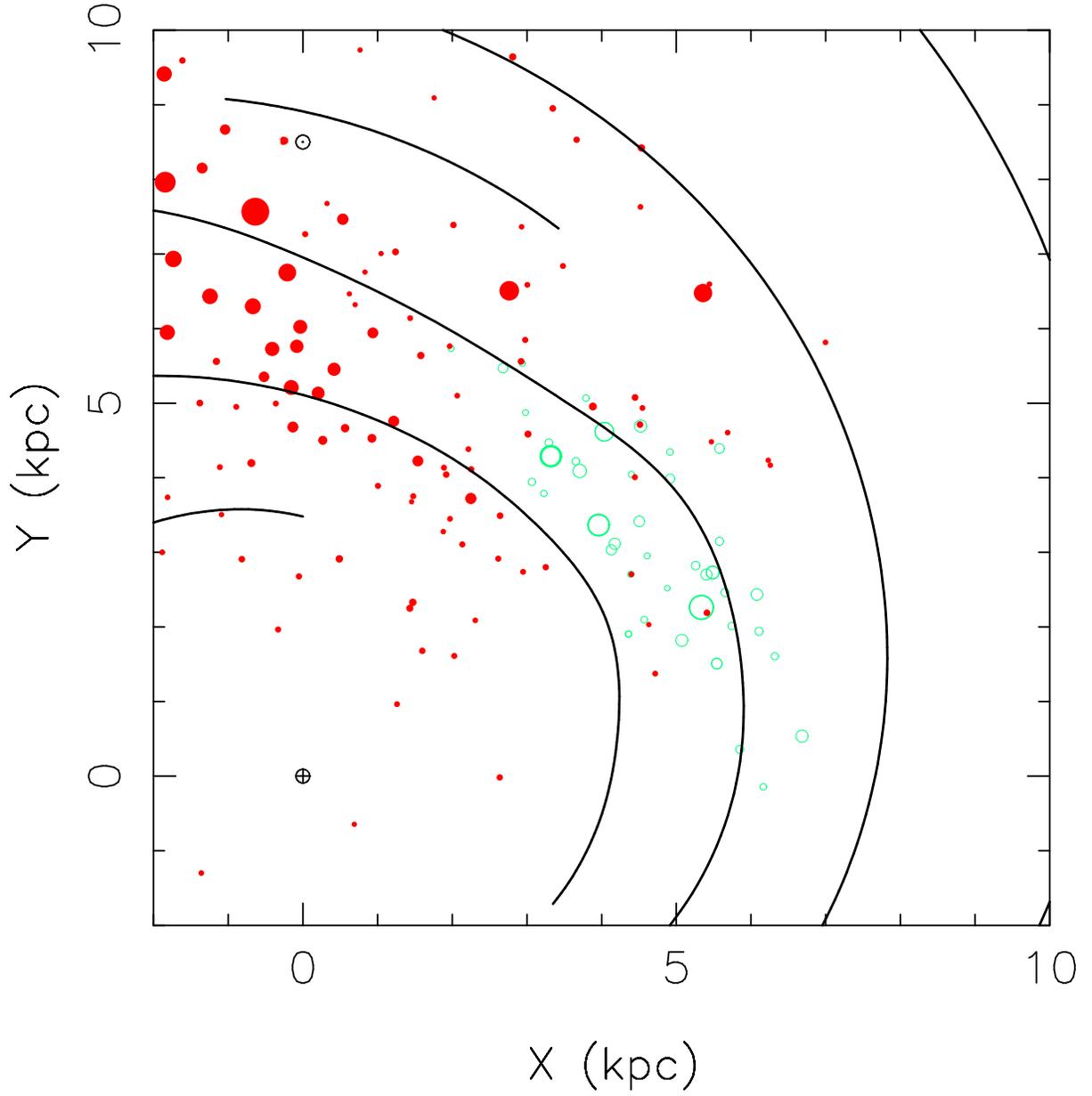}
\caption{Similar to the plot in Figure~\ref{fig:xyplot}, except that
the quantity plotted is the departure of the measured pulse-broadening
time (\taud) from the prediction of the NE2001 model (\taudnew);
the size of the symbol is proportional to the absolute value
of log(\taud/\taudnew). As for Figure~\ref{fig:xyplot}, the filled
circles represent the published data, while the open circles are the
new measurements. }
\label{fig:xyplotzoom}
\end{figure*}

Figure~\ref{fig:xyplot} shows the distribution of inferred SMs at
the locations of pulsars. The spiral arm locations are adopted from
the NE2001 model and the pulsar distances are the revised estimates
using this new electron density model. A more useful quantity is
the departure of the measured quantity (\taud\ or SM) from the
model predictions. In order to examine this in detail, we plot the
quantity $\vert \log(\taud/\taudnew)\vert$ at the locations of the
pulsars (Fig.~\ref{fig:xyplotzoom}).  Significant departures are seen
toward many LOSs. In the case of low-DM pulsars, these may be due
largely to measurement errors due to refractive scintillation effects
\citep{bhat1999a}. For distant, high-DM pulsars, departures from the
model predictions are in general larger in the inter-arm region. 
Most published data are in good agreement with the model predictions 
as expected, while several of the new measurements differ significantly 
from the model expectations.

\begin{figure*}
\epsscale{1.5}
\plotone{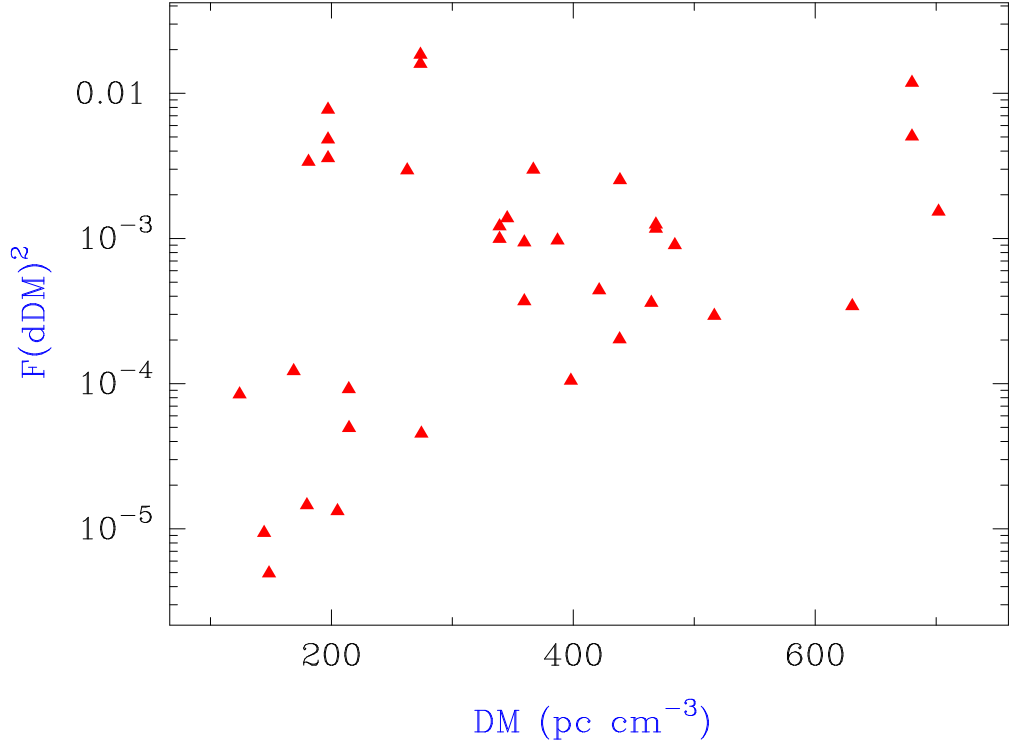}
\caption{Estimates of $F_c(\delta \mbox{DM})^2$ for the clumps of
enhanced scattering, derived from the {\em excess\/} scattering measures
(Table~\ref{tab:sm}), are plotted against the respective pulsar dispersion
measures.  The results are for a clump size of $\sim 10$\,pc and a volume
number density $\sim 1$\,\ncu\ for the clumps. For a fluctuation parameter
of $F_c = 10$, these results imply excess DM within the range $7 \times
10^{-4}$--$4 \times 10^{-2}$\,\dmu. }
\label{fig:fdm}
\end{figure*}

A closer examination of Figures~\ref{fig:tau-ne2001} and \ref{fig:tau-tc93} 
reveals that despite the general agreement seen with a large number of measurements,
the models still underestimate the total amount of scattering for many
LOSs.  The underestimates are accounted for easily by relaxing one or more
assumptions that underly the calculation of SM and its interpretation,
as has been pointed out by \citet{CL2002b}.  In particular, clumps of
enhanced scattering are likely due to unmodeled features associated
with \HII\ regions or supernova shocks.  Following \citet{CL2002b}
(see also \citet{chatterjeeetal2001}), we characterize such ``clumps''
in terms of the incremental SM and DM due to them.  We calculate the
increments associated with a clump as
\be
\delta \DM = \nec  \delta s
\label{eq:ddm}
\ee
and
\be
\delta \SM = \cnc \delta s,
\label{eq:dsm}
\ee
where \nec is the mean electron density and \cnc is a measure of the
fluctuating electron density 
inside a clump and $\delta s$ is the size of the clump region. The 
parameter \cnc can be expressed in terms of the electron density and 
the ``fluctuation parameter,'' \fc\ (see TC93; \citet{CL2002a}),
\be
\cnc = \csm \fc\ \necsq 
\label{eq:cnc}
\ee
where \csm is a numerical constant that depends on the slope of
the wavenumber spectrum, and is defined as 
$ \csm = \left [ 3 (2 \pi )^{1/3} \right] ^{-1} \ku $ for a
Kolmogorov spectrum, 
where the scale factor \ku = 10.2 \kuu yields SM in the conventional
units of \smu.  The fluctuation parameter \fc\ depends on the outer scale
(\lo), filling factor ($\eta$), and the fractional rms electron density
inside the clump. It is defined as (TC93)
\be 
\fc\ = \zeta \epsilon^2 \eta ^{-1} \lo ^{-2/3},
\ee
where $\zeta = \langle \overline{\nele ^2} \rangle / \langle
{\overline{\nele}} \rangle ^2 $, and $\epsilon = \langle (
\delne )^2 \rangle / \overline{ \nele } ^2$. From equations
\ref{eq:ddm}--\ref{eq:cnc}, the ratio of increments in SM and DM is
given by
\be
{ \delta \SM \over \delta \DM} = \csm \fc \nec.
\ee
The above expression can be re-written as 
\be 
\delta \SM  = \csm { \fc\ (\delta \DM)^2 \over \delta s}.
\ee
For large distances of a few to several kpc that are relevant for our
measurements, it is a fair assumption that the LOS to the pulsar may
encounter several such clumps.  Assuming a clump thickness of $\sim$10 pc
(typical size of known \HII\ regions), and a volume number density for
clumps, \nc\ $\sim$1 \ncu, we obtain the values of $\fc(\delta \DM)^2$
for the subset of measurements in Table~\ref{tab:sm} that show excess
SM (see also Fig.~\ref{fig:fdm}). The constraints derived from our data
lie within a broad range of
\be
10^{-5.3} < F_c (\delta \DM)^2 < 10^{-1.8}, 
\ee
which is consistent with values needed to account for the excess
scattering toward the LOS to pulsar B0919+06 derived by
\citet{chatterjeeetal2001}.  If we assume a fluctuation parameter
(\fc) of 10 for the clumps, which is consistent with values in TC93
and \citet{CL2002a}, the required range in $\delta\DM$ is $ 7 \times
10^{-4} < \delDM < 4 \times 10^{-2} $ \dmu.  For the assumed size of 10
pc for the clumps, this implies $10^{-5} \la \nec \la 4 \times 10^{-3}
$ \neu.  In reality, both the fluctuation parameter \fc, as well as
the sizes ($\delta s$) and number of clumps (\nc), will vary with the
LOS; nonetheless, the inferred values of $\delta$SM are such that the
derived constraints on the clumps are well within the range of physical
possibilities.  Note also that the implied perturbations of \DM\
are rather small, a fact that highlights the situation that relatively
small changes in the local mean electron density can translate into
large changes in the amount of scattering.

\section{Implications for the Electron Density Wavenumber Spectrum} 
\label{s:impli}

Our measurements of the scaling index $\alpha$ and the implied power
spectral slopes $\beta$ are summarized in Table~\ref{tab:scaling}.
In a few cases, the estimates of $\alpha$ are consistent with the
simple, Kolmogorov scaling of $\alpha=4.4$ (e.g. PSRs~J1853+0545,
J1913+1145, and J1920+1110). However, in most cases the measured scaling
is significantly weaker than even $\nu^{-4}$ (e.g. PSR~J1856+0404).
Overall the measurements show a possible departure from the traditional
expectation of $\nu^{-4.4}$ scaling, with a mean scaling index $\meanalpha
\approx 3.12 \pm 0.13 $ using the results for PBF$_1$, and $\approx
3.83 \pm 0.19 $ using those for PBF$_2$, in agreement with other recent
work \citep{lohmer2001}, and also comparable to a global scaling index
$3.86 \pm 0.16$ inferred from our parabolic fit to \taud\ vs DM data. 

\begin{figure*}
\epsscale{1.45}
\plotone{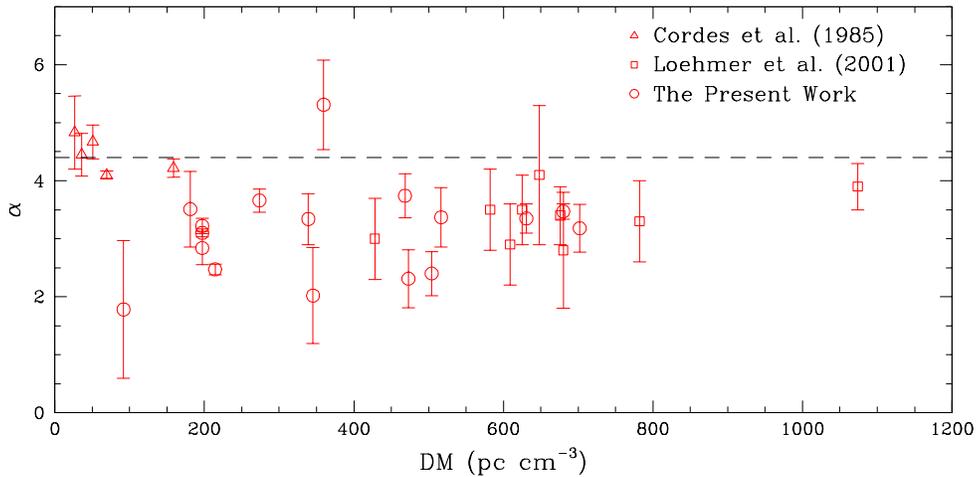}
\caption{Measurements of frequency scaling index ($\alpha_1$) against the
respective DMs. The results for low-DM objects \citep{CWB1985} are derived
from measurements of decorrelation bandwidths. For PSR~J1852+0031, the
only object common between our sample and that of \citet{lohmer2001},
estimates of $\alpha$ are consistent within measurement errors. The
dashed line corresponds to the Kolmogorov scaling index. }
\label{fig:alpha}
\end{figure*}

Figure~\ref{fig:alpha} summarizes the current state of the estimates
of $\alpha$ derived from measurements of decorrelation bandwidths and
pulse-broadening times. In addition to the present work which yielded 
$\alpha$ for 15 objects, this includes the recent measurements from 
\citet{lohmer2001} (for 9 high-DM objects) and those from \citet{CWB1985} 
(for 5 objects at low DMs) derived from measurements of \nd. Barring a 
few outlier cases, it appears that the scaling index is lower for objects 
of larger DMs ($\ga$ 200 \dmu), while it seems consistent with the 
Kolmogorov expectation for objects at lower DMs (although these are
only 5 in number). A similar result is also indicated by our analysis
of the DM dependence of pulse broadening times (\S~\ref{s:inner}).

We now return our attention to the dependence of the scaling index on the
PBF form adopted for the analysis.  PSR~J1853+0545 is a particularly
illustrative example.  For this object, based on the results for PBF$_1$,
we estimate a mean scaling index much lower than 4.4, $\meanalpha =
3.1 \pm 0.2$.  However, use of PBF$_2$ yields scaling indices that
agree well with that expected for a $\beta=11/3$ spectrum.  Naturally,
the two cases may lead to widely different interpretations in terms of
the nature of the wavenumber spectrum. Similarly, for PSR~J1857+0526,
while the estimate of $\alpha$ deduced from \taud\ values obtained for the
case of PBF$_1$ imply a power-law index that approaches the Kolmogorov
value, even if it is a little low, the results for the case of PBF$_2$
yield a much lower value.

Given all of this, it is important to attempt to use an approximately
correct form for the PBF before attempting any serious interpretation in
terms of the nature of the spectrum.  A mere departure from the expected
$\nu ^{-4.4}$ scaling need not necessarily signify an anomaly for the
scattering along that LOS.  However, in general it is clearly difficult
to know what is the correct form of the PBF for a given LOS.  
Additional figures of merit
such as the derived intrinsic pulse shapes, in particular their dependence
with frequency, and the number of CLEAN components \citep{bhatetal2003}
may help to resolve ambiguities in some cases, but the general problem
remains a difficult one.

\medskip
\medskip
\medskip
\medskip
\section{Summary and Conclusions}
\label{s:summary}

We have used multi-frequency radio data obtained with the Arecibo
telescope for a sample of 98, mostly distant, high-DM, pulsars to
measure in particular the pulse-broadening effect due to propagation
in the inhomogeneous interstellar medium. For 81 of these objects
we obtained data at 0.4, 1.2, 1.5 and 2.4 GHz, while data for the
remaining 17, at 0.4, 1.2 and 1.5 GHz, are from the recent work
of \citet{lorimeretal2002}.  We employed a CLEAN-based deconvolution
method to measure pulse-broadening times.  In this process we tested
two possible forms of the pulse-broadening function that characterizes
scattering along the LOS.  As a by-product, the method also yields
estimated shapes of the intrinsic pulse profiles.

The present work has resulted in new measurements of pulse-broadening
time for 56 pulsars, and upper limits for 31 pulsars.
These data, along with similar measurements from other published
work, were compared with the predictions from models for the Galactic
free electron density. New measurements allow an independent test of
the electron density model recently developed by \citet{CL2002a}. While
a majority of the data is in reasonable agreement with the model 
predictions, evidence for excess scattering is seen for many LOSs. We 
consider the possibility whereby the excess scattering can be accounted 
for by using ``clumps,'' regions of enhanced scattering in the Galaxy. 
Depending on the distance, a given LOS may contain one or more of such 
clumps, and we derive useful constraints on their properties.

For a small subset of objects, our data also allow estimation of the
frequency scaling indices for the pulse-broadening times, most of which
show significant departures from the traditional $\nu^{-4.4}$ behavior
expected for the case of a Kolmogorov power-law form for the spectrum
of density irregularities. Our analysis also suggests that the inferred
scaling indices depend on the type of PBF adopted for the analysis. We
combined our data with those from published work to revise the empirical
relation connecting pulse-broadening times and dispersion measures. The
inferred frequency scaling index from a globat fit is 3.9$\pm$0.2, less 
than that expected for the case of a Kolmogorov spectrum.  Our analysis 
also suggests the possibility of an inner scale in the range 
$\sim$300--800 km for the spectrum of turbulence. Further, the 
intrinsic pulse shapes deduced from our analysis for several of 
the pulsars are likely to be comparable to the actual pulse shapes, 
and hence may prove useful for applications such as the study of 
pulsar emission properties.

\bigskip
We thank Bill Sisk, Jeff Hagen and Andy Dowd for developing the WAPP
system at the Arecibo Observatory, which was crucial for providing much
of the data analyzed in this paper.  This work was supported by NSF grants
AST9819931, AST0138263 and  AST0206036  to Cornell University, AST0205853
to Columbia University, and AST0206205 to Princeton University.  NDRB is
supported by an MIT-CfA Postdoctoral Fellowship at Haystack Observatory.
DRL is a University Research Fellow funded by the Royal Society.  Arecibo
Observatory is operated by the National Astronomy and Ionosphere Center,
which is operated by Cornell University under cooperative agreement with
the National Science Foundation (NSF).



\clearpage

\twocolumn[
\begin{center} {\sc APPENDIX} \end{center}
\appendix
\section{\sc profile database}

\begin{figure*}
\epsscale{1.8}
\plotone{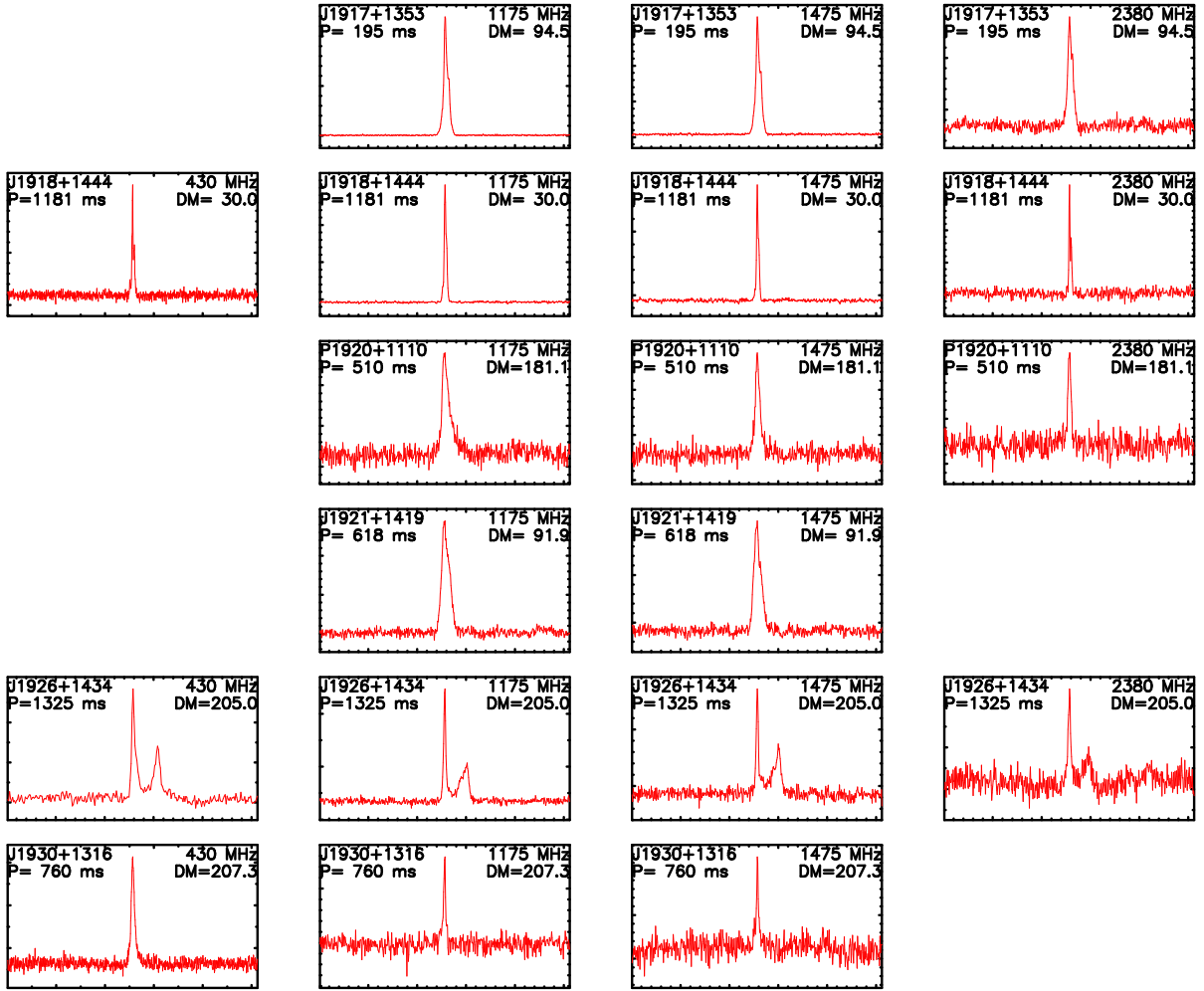}
\caption{Integrated pulse profiles of pulsars from Arecibo observations
at 430, 1175, 1475 and 2380 MHz. Data at 430 MHz were taken with the PSPM,
and those at higher frequencies were taken with the WAPP. All profiles are
plotted with a pulse phase resolution of 2 milli-periods, where consistent
with the data acquisition time resolution (Table~\ref{tab:obs}). The
highest point in the profile is placed at phase 0.5. The pulsar ID,
period and the dispersion measure are indicated at the top of each panel,
along with the center frequency of observation. Objects with labels
(top left) starting with `P' refer to new discoveries from the Parkes
multibeam survey, and those with `J' are previously known pulsars. }
\label{fig:profs}
\end{figure*}

The basic data (i.e. the pulse profiles in Fig.~\ref{fig:profs}) 
presented in this paper are also available as an electronic data
set. The full database is packaged as a {\tt gzipped tar} file, 
{\tt AOmultifreq\_profs.tar.gz} (which includes 345 pulse profiles from 
our observations), and is available over the Internet 
{\tt http://web.haystack.mit.edu/staff/rbhat/aoprofs} or can be 
downloaded via anonymous {\tt ftp} from the ftp site 
{\tt web.haystack.mit.edu} 
(the directory is {\tt pub/rbhat/aoprofs}). These profiles are 
stored as individual files in simple ascii format, which consists 
of a header line of the basic observing parameters followed by an 
ascii list of pulse bin number and the intensity value (in 
arbitrary units) in a two-column format. Each file is given a 
generic name of the format {\tt pulsar.freq.prf}, where {\tt pulsar} 
is the name of the pulsar and {\tt freq} is the frequency of 
observation in MHz. An example header is shown below, along 
with a description of the various parameters included in the 
header.

\medskip
{\tt \# mjdobs mjdsec per np freq refdm nbin siteid scanid source}
\medskip

where
\medskip

\begin{tabular}{lll}
{\tt mjdobs} &:& Date of observation (MJD) \\
{\tt mjdsec} &:& Time of observation (seconds, UTC) with \\
             & & respect to {\tt mjdobs} \\
{\tt per   } &:& Pulse period (seconds) \\
{\tt np    } &:& Pulse count \\
{\tt freq  } &:& Frequency of observation (MHz) \\
{\tt refdm } &:& Dispersion measure (\dmu) \\
{\tt nbin  } &:& Number of bins in the pulse profile \\
{\tt siteid} &:& Site ID of observations (`3' for Arecibo) \\
{\tt scanid} &:& Scan number of observation \\
{\tt source} &:& Source name \\
\end{tabular}
]

%
\def\tnm{\tablenotemark}
\begin{deluxetable}{cccccc}
\tablecolumns{6} 
\tablewidth{0pc} 
\tablecaption{Data acquisition parameters.\label{tab:obs}} 
\tablehead{ 
\colhead{Frequency} & 
\colhead{Bandwidth} & 
\colhead{Time} & 
\colhead{Spectral} & 
\colhead{Instrument} & 
\colhead{Integration Time\tnm{a}} \\    
\colhead{(MHz)} & 
\colhead{(MHz)} & 
\colhead{Resolution} & 
\colhead{Channels} & 
\colhead{} & 
\colhead{per Scan (min)} \\ 
}
\startdata 

\phn 430 	 & \phn \phn 8 	 & $10^{-3}\,P$	 & 128	 & PSPM & 10 \\
1175	 & 100	 & 256$\,\mu$s	 & 256	 & WAPP	 & \phn 5  \\
1475	 & 100	 & 256$\,\mu$s	 & 256	 & WAPP	 & \phn 5  \\
2380	 & \phn 50	 & 256$\,\mu$s	 & \phn 64	 & WAPP	 & 10 \\
\enddata 
\tablenotetext{a}{Multiple scans were made for pulsars with low signal-to-noise ratio in the first pass.}

\end{deluxetable} 


\clearpage
\def\tnm{\tablenotemark}
\begin{deluxetable}{ccccccccccccc}
\tabletypesize{\tiny}
\tablecolumns{10} 
\tablewidth{0pc} 
\tablecaption{New measurements of pulse-broadening times and predictions from the electron density models.\label{tab:meas}} 
\tablehead{ 
\colhead{} & \colhead{} & \colhead{} & \colhead{} & 
\colhead{} & \colhead{} & \colhead{} & 
\colhead{} & \colhead{} & \colhead{} & \colhead{}  \\
\colhead{PSR} & \colhead{Ref\tnm{a}} & \colhead{Period} & \colhead{DM} & 
\colhead{$l$} & \colhead{$b$} & \colhead{$\nu$} & 
\multicolumn{2}{c}{${\rm PBF_1}$} & \multicolumn{2}{c}{${\rm PBF_2}$} & 
\colhead{$\tau _{d,tc93}$} & \colhead{$\tau _{d,ne2001}$}  \\
\colhead{}    & \colhead{}  & \colhead{(ms)} & \colhead{(\dmu)} & 
\colhead{($^{\circ}$)} & \colhead{($^{\circ}$)}   & \colhead{(MHz)} & 
\colhead{$\taud$(ms)}   & $f_r$  & \colhead{$\taud$(ms)}   & $f_r$ & \colhead{(ms)} & \colhead{(ms)} \\
 & & &  &  &  &  & &  &  & &  &  \\
\colhead{(1)} & \colhead{(2)} & \colhead{(3)} & \colhead{(4)} & 
\colhead{(5)} & \colhead{(6)} & \colhead{(7)} & 
\colhead{(8)} & \colhead{(9)} & \colhead{(10)} & \colhead{(11)} & 
\colhead{(12)} & \colhead{(13)}  \\
}
\startdata 
J1848+0826 & 1,400 &  328.64 &  90.8  & 40.1    &   4.6 & 1175  & $<$0.2          &    & $<$0.2         &    & 0.003 &    0.001  \\ 
J1849+0127 & 2,1400&  542.11 &  214.4 &  33.9   &   1.2 &   430 & 78$\pm$21       &0.9 & $\cdots^b$     &    & 13.4  &    18.42  \\
           &       &         &        &         &       & 1175  & 6.5$\pm$2.1     &1.2 & 3.3$\pm$1      &0.1~& 0.161 &    0.221  \\ 
           &       &         &        &         &       & 1475  & $<$3.4          &    & $<$1.7         &    & 0.059 &    0.081  \\ 
J1850+0026 & 2,1400&1081.76  & 197.2  & 33.2    &  0.5  &  430  & 9.6$\pm$2.4     &2.5 & 4.8$\pm$1.1    &0.2~& 10.56 &    18.97  \\
           &       &         &        &         &       & 1175  & $<$4.2          &    & $<$1.1         &    & 0.127 &    0.228  \\ 
	   &       &         &        &         &       &       &                 &    &                &    &       &           \\
J1851+0118 & 3,1400& 907.05  & 413.0  & 34.1    &  0.7  &  1175 & $<$5            &    & $<$3           &    & 1.027 &    2.917  \\ 
J1851+0418 & 1,300 & 284.71  & 112.0  & 36.7    &  2.1  &  1175 & $<$3.5          &    & $<$1           &    & 0.010 &    0.008  \\ 
J1852+0031 & 1,1400& 2180.06 &  680.0 &  33.5   &   0.1 &  1175 &  495  $\pm$25   &2.5 &271   $\pm$5.7  &7.9~& 4.846 &    411.6  \\ 
           &       &         &        &         &       &  1475 &  225$\pm$14     &1.6 &127   $\pm$3.7  &1.8~& 1.782 &    151.3  \\ 
J1852+0305 & 2,1400&1326.06  & 315.6  & 35.7    &  1.3  &  1175 &  $<$14          &    & $\cdots^b$     &    & 0.448 &    0.728  \\ 
	   &       &         &        &         &       &       &                 &    &                &    &       &           \\
J1853+0056 & 2,1400& 275.56  & 182.4  & 33.9    &  0.1  &  1175 &  $<$2           &    & $<$1           &    & 0.089 &    0.196  \\ 
J1853+0505 & 3,1400&  905.21 &  273.6 &  37.6   &   2.0 &  1175 &  124  $\pm$14   &0.5 & $\cdots^b$     &    & 0.243 &    0.193  \\ 
           &       &         &        &         &       &  1475 &  54    $\pm$3   &1.4 & 23$\pm$4       &1.8~& 0.089 &    0.071  \\ 
J1853+0545 & 5,1400&  126.39 &  198.7 &  38.2   &   2.3 &  1175 &  13.6  $\pm$2   &0.4 & 8.2  $\pm$0.3  &2.5~& 0.098 &    0.049  \\ 
           &       &         &        &         &       &  1475 &  7.1   $\pm$0.9 &0.3 & 3    $\pm$0.5  &0.8~& 0.036 &    0.018  \\ 
           &       &         &        &         &       &  2380 &  1.5   $\pm$0.2 &0.1 & 0.4  $\pm$0.1  &1.7~& 0.004 &    0.002  \\ 
	   &       &         &        &         &       &       &                 &    &                &    &       &           \\
J1855+0422 & 2,1400& 1677.98 &  438.6 &  37.2   &   1.2 &  1175 &  27    $\pm$3   &0.2 & 16.6 $\pm$2.5  &0.9~& 1.004 &    1.803  \\ 
           &       &         &        &         &       & 1475  & $<$12           &    & $<$4           &    & 0.369 &    0.663  \\ 
J1856+0113 & 1,1400& 267.46  &  96.7  & 34.5    & --0.5 & 1175  & $<$1            &    & $<$1           &    & 0.006 &    0.007  \\ 
J1856+0404 & 2,1400&  420.22 &  345.3 &  37.1   &   0.8 &  1175 &  9.5   $\pm$4   &0.3 & 4.8  $\pm$1.6  &0.04& 0.709 &    1.279  \\ 
           &       &         &        &         &       & 1475  & 6$\pm$3         &0.2 & 2.8$\pm$1      &0.2~& 0.261 &    0.470  \\ 
	   &       &         &        &         &       &       &                 &    &                &    &       &           \\
J1857+0057 & 1,300 & 356.93  &  83.0  & 34.4    & --0.8 &  430  & $<$5            &    & $<$1           &    & 0.271 &    0.229  \\ 
J1857+0210 & 2,1400&  630.94 &  783.2 &  35.5   &  --0.3&  1175 &  13.4  $\pm$3.6 &0.06& 6.1  $\pm$0.65 &0.04& 9.004 &    19.96  \\ 
           &       &         &        &         &       & 1475  &  $<$5.5         &    & $<$4           &    & 3.311 &    7.340  \\ 
J1857+0212 & 1,1400&  415.80 &  504.0 &  35.5   &  --0.2&  1175 &  3.8   $\pm$0.9 &0.1 & 1.2  $\pm$0.3  &1.8~& 2.749 &    6.135  \\ 
           &       &         &        &         &       & 1475  &  2.2$\pm$0.3    &0.8 & 0.5$\pm$0.4    &0.7~& 1.011 &    2.256  \\ 
	   &       &         &        &         &       &       &                 &    &                &    &       &           \\
J1857+0526 & 5,1400&  349.92 &  468.3 &  38.4   &   1.2 &  1175 &  14.5  $\pm$1.7 &0.4 & 6    $\pm$1    &0.6~& 0.975 &    1.700  \\ 
           &       &         &        &         &       & 1475  &  6.2$\pm$1.3    &0.04& 3$\pm$1        &0.03& 0.359 &    0.625  \\ 
J1857+0809 & 3,1400& 502.96  & 284.2  & 40.8    &  2.5  &  1175 &  $<$3.5         &    & $<$2           &    & 0.195 &    0.061  \\ 
J1857+0943 & 1,400 &   5.36  &  13.3  & 42.2    &  3.2  &  430  &  $\cdots^c$     &    & $\cdots^c$     &    & 0.001 &    0.001  \\ 
J1858+0215 & 2,1400&  745.77 &  702.0 &  35.7   &  --0.4&  1175 &  38    $\pm$3   &3.3 &22.5  $\pm$6.5  &0.7~& 6.267 &    13.77  \\ 
           &       &         &        &         &       & 1475  &  18.4$\pm$4.7   &0.4 & 11.7$\pm$4.8   &0.05& 2.304 &    5.062  \\ 
	   &       &         &        &         &       &       &                 &    &                &    &       &           \\
J1858+0241 & 3,1400&4693.60  & 341.7  & 36.1    & --0.2 &  1175 & $<$22           &    & $<$16          &    & 0.758 &    1.702  \\ 
J1900+0227 & 2,1400& 374.24  & 201.1  & 36.1    & --0.8 &  1175 & $<$4            &    & $<$2           &    & 0.116 &    0.178  \\ 
J1901+0156 & 1,400 & 288.22  & 102.1  & 35.7    & --1.2 &  430  & 3.5$\pm$1       &0.2 & 0.7$\pm$0.3    &0.2~& 0.619 &    0.164  \\ 
           &       &         &        &         &       & 1175  & $<$2            &    & $<$1           &    & 0.007 &    0.002  \\ 
J1901+0331 & 1,400 &  655.48 &  401.2 &  37.2   &  --0.5&   430 &  60    $\pm$3   &0.8 & 44   $\pm$4.2  &1.5~& 107.9 &    76.16  \\ 
           &       &         &        &         &       & 1175  &  $<$3           &    & $<$1           &    & 1.295 &    0.914  \\ 
	   &       &         &        &         &       &       &                 &    &                &    &       &           \\
J1901+0355 & 3,1400& 554.80  & 546.2  & 37.5    & --0.3 &  1175 &  $<$4           &    & $\cdots^b$     &    & 2.97  &    6.459  \\ 
J1901+0413 & 2,1400& 2662.88 &  367.0 &  37.8   &  --0.2&   430 &  $<$558         &    & $\cdots^b$     &    & 79.5  &    161.7  \\
           &       &         &        &         &       & 1175  &  $<$13          &    & $<$3           &    & 0.954 &    1.940  \\ 
J1901+0716 & 1,1400&  644.02 &  252.8 &  40.5   &   1.2 &   430 &  10.1  $\pm$2.4 &0.07& 8.5$\pm$3.4    &0.04& 17.12 &    23.05  \\ 
           &       &         &        &         &       & 1175  &  $<$2.7         &    &  $<$1          &    & 0.205 &    0.277  \\ 
	   &       &         &        &         &       &       &                 &    &                &    &       &           \\
J1901+1306 & 1,400 &1830.72  &  75.0  & 45.7    &  3.9  & 1175  &  $\cdots^d$     &    &  $\cdots^d$    &    & 0.001 &    $\ll$1 \\ 
J1902+0556 & 1,400 &  746.60 &  179.7 &  39.4   &   0.4 &   430 &  12    $\pm$1.1 &1.1 & 5.7  $\pm$1.4  &0.3~& 4.674 &    10.05  \\ 
           &       &         &        &         &       & 1175  &  $<$4.2         &    &  $<$1.6        &    & 0.056 &    0.121  \\ 
J1902+0723 & 1,400 & 487.82  & 105.0  & 40.7    &  1.0  &  430  &  $<$8           &    &  $<$3          &    & 0.455 &    0.473  \\ 
J1903+0135 & 1,400 &  729.33 &  246.4 &  35.7   &  --1.8&   430 &  11.4  $\pm$0.9 &3.1~& 5.2  $\pm$0.6  &3.5~& 16.83 &    13.73  \\ 
           &       &         &        &         &       & 1175  &  $<$3           &    & $<$1           &    & 0.202 &    0.165  \\ 
	   &       &         &        &         &       &       &                 &    &                &    &       &           \\
J1903+0601 & 3,1400&  374.11 &  398.0 &  39.7   &   0.2 &  1175 &  2.7   $\pm$0.5 &0.03& 1   $\pm$0.4   &0.01& 1.100 &    2.093  \\ 
           &       &         &        &         &       & 1475  &  $<$1.7         &    & $<$1           &    & 0.405 &    0.77   \\ 
J1904+0004 & 1,400 &  139.53 &  233.7 &  34.4   &  --2.8&   430 &  3.1   $\pm$1   &0.7 & 4    $\pm$1    &0.8~& 10.78 &    4.66   \\ 
           &       &         &        &         &       & 1175  &  $<$1.4         &    & $<$0.7         &    & 0.129 &    0.56   \\ 
J1904+0412 & 2,1400&  71.09  & 185.9  & 38.1    & --0.9 &  430  &  $\cdots^d$     &    & $\cdots^d$     &    & 5.963 &    10.24  \\ 
	   &       &         &        &         &       &       &                 &    &                &    &       &           \\
J1904+0800 & 5,1400& 263.37  & 438.3  & 41.5    &  0.9  &  430  &  $\cdots^e$     &    & $\cdots^e$     &    & 61.02 &    118.7  \\
           &       &         &        &         &       &  1175 &  3     $\pm$0.3 &0.1 & 1.2  $\pm$0.12 &0.4~& 0.723 &    1.390  \\ 
           &       &         &        &         &       & 1475  &  $<$2           &    &  $<$1          &    & 0.266 &    0.511  \\ 
J1904+1011 & 1,400 &1856.64  & 135.0  & 43.4    &  1.9  &  430  &  $\cdots^b$     &    &  $<$4.4        &    & 1.015 &    0.937  \\
J1905+0400 & 3,1400&   3.78  &  25.8  & 38.0    & --1.2 &  430  &  $\cdots^d$     &    & $\cdots^d$     &    & 0.005 &    0.001  \\ 
	   &       &         &        &         &       &       &                 &    &                &    &       &           \\
J1905+0616 & 2,1400&  989.64 &  262.7 &  40.1   &  --0.2&  1175 &  13.5  $\pm$3.1 &0.04& 7.8  $\pm$2.1  &0.2~& 0.244 &    0.514  \\ 
           &       &         &        &         &       & 1475  &  $<$5.8         &    &  $<$1.5        &    & 0.09  &    0.189  \\ 
J1905+0709 & 1,1400& 648.05  & 269.0  & 40.8    &  0.3  &  430  &  $\cdots^b$     &    & 41$\pm$10      &0.1~& 20.95 &    43.77  \\ 
           &       &         &        &         &       & 1175  & 7  $\pm$4       &0.03& 3.2$\pm$1.6    &0.02& 0.251 &    0.525  \\ 
J1906+0641 & 1,1400&  267.29 &  473.0 &  40.5   &  --0.2&   1175&  4.4   $\pm$1.1 &0.2 & 2.4$\pm$0.4    &0.7~& 1.347 &    111.1  \\ 
           &       &         &        &         &       & 1475  &  2.6$\pm$0.7    &0.05& 1.1$\pm$0.4    &0.12& 0.495 &    40.85  \\ 
	   &       &         &        &         &       &       &                 &    &                &    &       &           \\
J1906+0912 & 2,1400& 775.29  & 260.5  & 42.8    &  1.0  &  1175 & $<$8            &    &  $<$4          &    & 0.189 &    0.278  \\ 
J1907+0534 & 2,1400&1138.31  & 526.7  & 39.7    & --0.9 &  1175 & $<$14           &    &  $<$7          &    & 1.351 &    2.674  \\ 
J1907+0740 & 2,1400&  574.65 &  329.3 &  41.5   &   0.1 &   430 &  10.1$\pm$3.8   &0.03& 6.6  $\pm$1.6  &0.03& 41.23 &    85.56  \\
           &       &         &        &         &       & 1175  &  $<$2.3         &    &  $<$1          &    & 0.495 &    1.026  \\ 
J1907+0918 & 1,1400& 226.12  & 358.0  & 43.0    &  0.8  &  1175 & $<$2.4          &    &  $<$1.2        &    & 0.471 &    0.766  \\ 
	   &       &         &        &         &       &       &                 &    &                &    &       &           \\
J1907+1247 & 1,400 & 827.11  & 257.0  & 46.1    &  2.4  &  430  & $\cdots^b$      &    &  $<$1          &    & 7.631 &    3.25   \\
J1908+0457 & 1,400 &  846.83 &  360.0 &  39.2   &  --1.4&   430 &  $\cdots^b$     &    &  18$\pm$7      &0.02& 42.15 &    57.1   \\ 
           &       &         &        &         &       & 1175  &  $<$3.6         &    &  $<$1.8        &    & 0.506 &    0.69   \\ 
J1908+0500 & 1,400 &  291.03 &  201.4 &  39.3   &  --1.4&   430 &  4     $\pm$0.7 &0.74& 2.1  $\pm$0.25 &0.15& 7.703 &    9.7    \\ 
           &       &         &        &         &       & 1175  &  $<$0.5         &    &  $<$0.3        &    & 0.092 &    0.12   \\ 
	   &       &         &        &         &       &       &                 &    &                &    &       &           \\
J1908+0734 & 1,400 & 212.34  &  11.1  & 41.6    & --0.2 & 1175  &  $\cdots^c$     &    &  $\cdots^c$    &    &$\ll$1 &    $\ll$1 \\ 
J1908+0839 & 2,1400&  185.38 &  516.6 &  42.5   &   0.3 &  1175 &  5.6   $\pm$1.3 &0.4 & 3.5  $\pm$0.8  &0.2~& 1.131 &    2.595  \\ 
           &       &         &        &         &       & 1475  &  2.6$\pm$0.6    &0.55& 1.4$\pm$0.2    &0.3~& 0.416 &    0.954  \\ 
J1908+0909 & 2,1400&  336.53 &  464.5 &  43.0   &   0.5 &  1175 &  4.9   $\pm$0.9 &1.14& 2.4  $\pm$0.5  &1~~~& 0.757 &    1.655  \\ 
           &       &         &        &         &       & 1475  &  $<$3           &    & $\cdots^b$     &    & 0.279 &    0.609  \\ 
	   &       &         &        &         &       &       &                 &    &                &    &       &           \\
J1908+0916 & 1,400 &  830.31 &  250.0 &  43.1   &   0.6 &  430  &  $\cdots^b$     &    & 12.7$\pm$2.1   &0.04& 12.87 &    25.13  \\ 
J1909+0007 & 1,400 &1016.96  & 112.9  & 35.0    & --3.9 &  430  &  $<$2           &    & $<$1           &    & 1.006 &    0.195  \\ 
J1909+0254 & 1,400 & 989.85  & 172.1  & 37.5    & --2.6 &  430  &  $<$2.7         &    & $<$1.1         &    & 5.03  &    1.995  \\ 
J1909+0616 & 2,1400& 251.98  & 348.6  & 40.5    & --1.0 &  430  &  $\cdots^d$     &    & $\cdots^d$     &    & 46.15 &    69.79  \\
           &       &         &        &         &       & 1175  &  $<$5           &    & $<$2           &    & 0.554 &    0.837  \\ 
	   &       &         &        &         &       &       &                 &    &                &    &       &           \\
J1909+0912 & 2,1400&  222.93 &  421.4 &  43.1   &   0.3 &  1175 &  5.1$\pm$1.2    &0.04& 2.2  $\pm$0.5  &0.03& 0.675 &    1.681  \\ 
           &       &         &        &         &       & 1475  &  $<$2.2         &    & $<$1.1         &    & 0.248 &    0.618  \\ 
J1909+1102 & 1,400 &  283.65 &  148.4 &  44.7   &   1.2 &   430 &  1.5   $\pm$0.1 &2.4 &  0.33 $\pm$0.03&5.2~& 1.328 &    1.076  \\ 
           &       &         &        &         &       & 1175  &  $<$1           &    &  $\cdots^b$    &    & 0.016 &    0.013  \\ 
J1909+1450 & 1,400 & 996.04  & 119.5  & 48.1    &  2.9  & 1175  &  $\cdots^d$     &    & $\cdots^d$     &    & 0.003 &    0.012  \\ 
	   &       &         &        &         &       &       &                 &    &                &    &       &           \\
J1910+0358 & 1,300 &2330.30  &  78.8  & 38.6    & --2.3 &  430  &  $<$ 4.7        &    & $<$5.7         &    & 0.14  &    0.114  \\ 
J1910+0534 & 2,1400&  452.83 &  484.0 &  40.0   &  --1.6&  1175 &  12.5  $\pm$3.8 &0.7~& 5.4$\pm$2      &0.1~& 0.676 &    0.773  \\ 
           &       &         &        &         &       & 1475  &  $<$6.2         &    & $<$2.2         &    & 0.249 &    0.284  \\ 
J1910+0714 & 1,400 & 2712.51 &  124.1 &  41.5   &  --0.8&   430 &  9.3$\pm$2.6    &0.01& $<$1.1         &    & 0.887 &    1.628  \\ 
J1910+1231 & 1,400 & 1441.81 &  274.4 &  46.2   &   1.6 &   430 &  24.6  $\pm$6.4 &0.06& 14.7$\pm$1.8   &0.2~& 10.59 &    8.278  \\ 
           &       &         &        &         &       & 1175  &  $<$3           &    & $<$1.5         &    & 0.127 &    0.099  \\ 
	   &       &         &        &         &       &       &                 &    &                &    &       &           \\
J1912+1036 & 1,400 & 409.38  & 147.0  & 44.7    &  0.3  &  430  &  4.2$\pm$2.7    &0.04&  2.9$\pm$1.5   &0.03& 1.317 &    1.933  \\
           &       &         &        &         &       & 1175  &  $<$3           &    & $<$2           &    & 0.013 &    0.023  \\
J1913+0446 & 5,1400&1616.09  & 109.1  & 39.7    & --2.6 & 1175  &  $<$2           &    & $<$1.4         &    & 0.008 &    0.005  \\ 
J1913+0832{\tnm{m}}                                                                                            
           & 2,1400&134.40   &359.5   &43.0     & --0.9 &  1175 &  7.7   $\pm$1.7 &0.7 & 5.1  $\pm$1.4  &0.1~& 0.466 &    0.734  \\ 
           &       &         &        &         &       & 1475  &  2.3$\pm$1      &0.03& $\cdots^b$     &    & 0.171 &    0.27   \\ 
	   &       &         &        &         &       &       &                 &    &                &    &       &           \\
J1913+0832{\tnm{i}}                                                                                            
           &       &         &        &         &       &  1175 &  6.6   $\pm$1.8 &0.12& 3  $\pm$1.4    &0.2~& 0.509 &    0.905  \\ 
           &       &         &        &         &       & 1475  &  $<$2           &    & $<$1           &    & 0.187 &    0.333  \\ 
J1913+0936 & 1,400 &1242.06  & 157.0  & 43.9    & --0.4 &  430  &  $<$1.3         &    & $<$0.4         &    & 1.874 &    35.38  \\
J1913+1000 & 3,1400& 837.21  & 419.4  & 44.3    & --0.2 &  1175 &  11.1$\pm$4     &0.03& 5$\pm$3        &0.03& 0.556 &    16.94  \\ 
           &       &         &        &         &       & 1475  &  $<$6           &    & $<$3           &    & 0.204 &    6.228  \\ 
	   &       &         &        &         &       &       &                 &    &                &    &       &           \\
J1913+1011 & 2,1400&  35.91  & 178.9  & 44.4    & --0.1 &  430  &  2$\pm$1        &0.3~& 0.9$\pm$0.7    &0.4~& 2.986 &    107.9  \\ 
           &       &         &        &         &       & 1175  &  $<$0.4         &    & $<$0.2         &    & 0.036 &    1.294  \\ 
J1913+1145 & 2,1400&  306.05 &  630.7 &  45.8   &   0.6 &  1175 &  9.2   $\pm$1   &0.3~& 5.6$\pm$1.3    &0.1~& 1.568 &    1.204  \\ 
           &       &         &        &         &       & 1475  &  4.3$\pm$0.5    &0.2 & 2.2$\pm$0.9    &0.01& 0.577 &    0.443  \\ 
J1913+1400 & 1,400 &  521.50 &  144.4 &  47.8   &   1.7 &   430 &  3.3   $\pm$0.6 &1.7 & $<$0.53        &    & 0.706 &    2.126  \\ 
           &       &         &        &         &       & 1175  &  $<$2           &    & $<$1           &    & 0.009 &    0.026  \\ 
	   &       &         &        &         &       &       &                 &    &                &    &       &           \\
J1914+1122 & 1,400 & 600.96  &  80.0  & 45.6    &  0.2  & 1175  &  $\cdots^d$     &    & $\cdots^d$     &    & 0.001 &    0.003  \\ 
J1915+0839 & 3,1400& 342.77  & 369.1  & 43.4    & --1.2 & 1175  &  $<$6           &    & $<$2           &    & 0.396 &    0.46   \\ 
J1915+0738 & 1,400 &1542.73  &  39.0  & 42.4    & --1.7 &  430  &  $\cdots^c$     &    &  $\cdots^c$    &    & 0.012 &    0.002  \\ 
J1915+1009 & 1,400 &  404.56 &  246.1 &  44.6   &  --0.6&   430 &  15.4  $\pm$1.3 &6.9 & 11.1 $\pm$1    &0.9~& 10.20 &    20.27  \\ 
           &       &         &        &         &       & 1175  &  $<$1.1         &    & $<$0.4         &    & 0.122 &    0.243  \\ 
	   &       &         &        &         &       &       &                 &    &                &    &       &           \\
J1915+1606 & 1,400 &   59.06 &  168.8 &  49.9   &   2.2 &  1175 &  0.33$\pm$0.1   &0.4 & $<$0.07        &    & 0.007 &    0.021  \\ 
J1916+0844 & 3,1400&  440.03 &  339.0 &  43.6   &  --1.1&  1175 &  7.7   $\pm$1   &0.1 & 4$\pm$0.6      &1.4~& 0.357 &    0.428  \\ 
           &       &         &        &         &       &  1475 &  3.6   $\pm$0.9 &0.1 & 1.1 $\pm$0.4   &0.3~& 0.131 &    0.158  \\ 
J1916+0951 & 1,400 & 270.26  &  61.4  & 44.5    & --0.9 &  430  &  1$\pm$0.4      &0.6 & $<$0.6         &    & 0.037 &    0.025  \\   
J1916+1030 & 1,400 &  628.92 &  387.0 &  45.1   &  --0.6&  1175 &  9.2$\pm$2.9    &0.02& 4.8  $\pm$1.5  &0.02& 0.4   &    0.714  \\   
           &       &         &        &         &       & 1475  &  $<$4           &    & $<$2           &    & 0.147 &    0.262  \\   
	   &       &         &        &         &       &       &                 &    &                &    &       &           \\
J1916+1312 & 1,400 &  281.86 &  236.9 &  47.5   &   0.7 &   430 &  2.8   $\pm$0.1 &5.7 & 0.9  $\pm$0.4  &2.4~& 5.958 &    12.93  \\   
           &       &         &        &         &       & 1175  &  $<$1           &    & $<$0.4         &    & 0.071 &    0.155  \\   
J1917+1353 & 1,300 & 194.62  &  94.5  & 48.2    &  0.8  & 1175  &  1.2$\pm$0.4    &0.3~& 0.4$\pm$0.1    &0.4~& 0.001 &    0.011  \\   
J1918+1444 & 1,400 &1181.08  &  30.0  & 49.0    &  0.9  &  430  & $\cdots^c$      &    & $\cdots^c$     &    & 0.006 &    0.001  \\   
J1918+1541 & 1,400 & 370.86  &  13.0  & 49.9    &  1.4  & 1175  & $\cdots^c$      &    & $\cdots^c$     &    &$\ll$1 &    $\ll$1 \\   
	   &       &         &        &         &       &       &                 &    &                &    &       &           \\
J1920+1110 & 2,1400&  509.85 &  181.1 &  46.1   &  --1.2&  1175 &  14    $\pm$3.3 &0.02& 7.4$\pm$1.4    &0.01& 0.032 &    0.06   \\ 
           &       &         &        &         &       & 1475  &  6.3$\pm$2.1    &0.4~& 2.9$\pm$1.7    &0.5~& 0.012 &    0.022  \\ 
J1921+1419 & 1,400 & 618.14  &  91.9  & 49.0    &  0.1  & 1175  &  6$\pm$4        &0.4 & 3.2$\pm$1.9    &0.9~& 0.002 &    0.01   \\   
           &       &         &        &         &       & 1475  &  4$\pm$3        &0.3 & 2.9$\pm$2.2    &0.05& 0.001 &    0.004  \\   
J1921+2003 & 4,400 & 760.70  & 101.0  & 54.1    &  2.8  &  430  &  $\cdots^b$     &    & $<$1.2         &    & 0.096 &    0.07   \\
	   &       &         &        &         &       &       &                 &    &                &    &       &           \\
J1923+1706 & 4,400 &  547.23 &  142.5 &  51.7   &   1.0 &   430 &  $\cdots^b$     &    & $<$1           &    & 0.406 &    1.23   \\
J1926+1434 & 1,400 & 1324.99 &  205.0 &  49.8   &  --0.8&   430 &  8.4$\pm$1.3    &4.2 & $<$7.4         &    & 1.934 &    5.43   \\   
J1926+1928 & 4,400 & 1346.05 &  445.0 &  54.1   &   1.5 &   430 &  $\cdots^b$     &    & 34.5 $\pm$6.8  &0.4~& 11.46 &    3.46   \\
J1927+1852 & 4,400 &  482.79 &  254.0 &  53.7   &   1.0 &   430 &  $\cdots^b$     &    & 5.8  $\pm$1.3  &0.5~& 2.754 &    1.97   \\
           &       &         &        &         &       &  1175 &  $\cdots^b$     &    & $<$1           &    & 0.033 &    0.024  \\
	   &       &         &        &         &       &       &                 &    &                &    &       &           \\
J1927+1856 & 4,400 &  298.34 &   90.0 &  53.8   &   1.0 &   430 & $\cdots^b$      &    & $<$0.2         &    & 0.09  &    0.09   \\
J1929+1844 & 4,400 &1220.38  & 109.0  & 53.8    &  0.5  & 1175  & $\cdots^b$      &    & $<$1.1         &    & 0.002 &    0.003  \\
J1930+1316 & 1,400 & 760.05  & 207.3  & 49.1    & --2.3 &  430  & $\cdots^b$      &    & 4.1$\pm$2.4    &0.1~& 1.526 &    1.89   \\  
           &       &         &        &         &       &  1175 &  $\cdots^b$     &    & $<$1.2         &    & 0.018 &    0.023  \\  
J1931+1536 & 4,400 &  314.37 &  140.0 &  51.3   &  --1.4&   430 &  $\cdots^b$     &    & 0.9 $\pm$0.2   &0.01& 0.33  &    0.87   \\
	   &       &         &        &         &       &       &                 &    &                &    &       &           \\
J1933+1304 & 4,400 &  928.38 &  177.9 &  49.3   &  --3.1&   430 & $\cdots^b$      &    & $<$0.25        &    & 0.660 &    0.334  \\
J1935+1745 & 4,400 &  654.44 &  214.6 &  53.6   &  --1.2&   430 &  15.9  $\pm$1.5 &1.2 & 9.2$\pm$3.3    &0.02& 0.751 &    2.72   \\
           &       &         &        &         &       & 1175  & $<$1.1          &    & $<$1.2         &    & 0.008 &    0.033  \\
J1942+1743 & 4,400 &  696.30 &  190.0 &  54.4   &  --2.7&   430 &  $\cdots$       &    &    5$\pm$1.5   &0.5~& 0.403 &    0.29   \\
J1944+1755 & 4,400 &1996.78  & 175.0  & 54.8    & --3.0 &  430  &  $\cdots^b$     &    &   $<$5         &    & 0.308 &    0.215  \\
	   &       &         &        &         &       &       &                 &    &                &    &       &           \\
J1945+1834 & 4,400 & 1068.80 &  215.0 &  55.5   &  --2.9&   430 &  $\cdots^b$     &    &3.3  $\pm$1.3   &0.2~& 0.7   &    0.524  \\
J2027+2146 & 4,400 & 398.20  &  96.8  & 63.5    & --9.5 &  430  &  $<$0.4         &    & $<$0.4         &    & 0.091 &    0.041  \\
\enddata 

\tablenotetext{a}{Reference(s) to pulsar parameters,
followed by the frequency band (in MHz) of the survey that
discovered the pulsar. (1) ATNF pulsar catalog, available at
http://www.atnf.csiro.au/research/pulsar/psrcat; (2) Morris et al. (2002);
(3) Hobbs et al. (2004), in preparation; (4) Lorimer et al. (2002);
(5) Kramer et al. (2003). Note that pulsars in (2), (3), and (5) (Parkes
multibeam survey discoveries), as well as in (4), are also available in
(1). }
\tablenotetext{b}{The PBF yields unphysical residuals from the
deconvolution for any realistic value of $\tau_d$ (see text).}
\tablenotetext{c}{Signal-to-noise ratio is too small to allow a meaningful fit to the PBF.}
\tablenotetext{d}{\taud\ is negligibly small.}
\tablenotetext{e}{The pulsar is not detected at this frequency.}
\tablenotetext{m,i}{Main pulse (m) and inter pulse (i) of the pulsar
(see also Fig.~\ref{fig:profs}).}

\end{deluxetable} 


\clearpage
\def\tnm{\tablenotemark}
\begin{deluxetable}{cccccccc}
\tablecolumns{8} 
\tablewidth{0pc} 
\tablecaption{Frequency scaling indices from measurements of pulse-broadening times.\label{tab:scaling}} 
\tablehead{ 
&&&\multicolumn{2}{c}{PBF$_1$ deconvolution} && 
   \multicolumn{2}{c}{PBF$_2$ deconvolution} \\
\cline{4-5}  \cline{7-8} 
\colhead{PSR} & \colhead{$\nu _1$} & \colhead{$\nu _2$} & \colhead{$\alpha_1$} &
\colhead{$\beta_1$} && \colhead{$\alpha_2$} & \colhead{$\beta_2$} \\
\colhead{}    & \colhead{(MHz)}  & \colhead{(MHz)} & \colhead{} & 
\colhead{} & \colhead{}   & \colhead{} \\
\colhead{(1)} & \colhead{(2)} & \colhead{(3)} & \colhead{(4)} & 
\colhead{(5)} && \colhead{(6)} & \colhead{(7)} \\
}
\startdata 

J1849+0127 &  430 & 1175 & 2.5  $\pm$ 0.1  &   $\cdots^a$                  &&    \nodata      &  \nodata          \\             
J1852+0031 & 1175 & 1475 & 3.5  $\pm$ 0.1  &   4.7  $\pm$   0.4  && 3.3  $\pm$ 0.1  &  5    $\pm$  0.2  \\
J1853+0505 & 1175 & 1475 & 3.7  $\pm$ 0.2  &   4.4  $\pm$   0.5  &&    \nodata      &  \nodata          \\
J1853+0545 & 1175 & 1475 & 2.8  $\pm$ 0.3  &   6.8  $\pm$   1.5  && 4.4  $\pm$ 0.3  &  3.7  $\pm$  0.5  \\
           & 1475 & 2380 & 3.2  $\pm$ 0.1  &   5.3  $\pm$   0.5  && 4.2  $\pm$ 0.2  &  3.8  $\pm$  0.3  \\
           & 1175 & 2380 & 3.1  $\pm$ 0.1  &   5.7  $\pm$   0.3  && 4.3  $\pm$ 0.1  &  3.8  $\pm$  0.1  \\
J1856+0404 & 1175 & 1475 & 2.0  $\pm$ 0.8  &   $\cdots^a$                  && 2.4  $\pm$ 0.7  &  $\cdots^a$                 \\
J1857+0212 & 1175 & 1475 & 2.4  $\pm$ 0.4  &   $\cdots^a$                  && 3.9  $\pm$ 1.4  &  4.1  $\pm$  3.2  \\
J1857+0526 & 1175 & 1475 & 3.7  $\pm$ 0.4  &   4.3  $\pm$   1    && 3    $\pm$ 0.6  &  6.1  $\pm$  2.6  \\
J1858+0215 & 1175 & 1475 & 3.2  $\pm$ 0.4  &   5.4  $\pm$   1.6  && 2.9  $\pm$ 0.8  &  6.6  $\pm$  3.8  \\
J1906+0641 & 1175 & 1475 & 2.3  $\pm$ 0.5  &   $\cdots^a$                  && 3.4  $\pm$ 0.6  &  4.8  $\pm$  2    \\
J1908+0839 & 1175 & 1475 & 3.4  $\pm$ 0.5  &   4.9  $\pm$   1.7  && 4.0  $\pm$ 0.4  &  4    $\pm$  1    \\
J1913+0832 & 1175 & 1475 & 5.3  $\pm$ 0.8  &   3.2  $\pm$   1.1  &&    \nodata      &    \nodata        \\
J1913+1145 & 1175 & 1475 & 3.4  $\pm$ 0.3  &   5    $\pm$   0.8  && 4.1  $\pm$ 0.8  &  3.9  $\pm$  1.6  \\
J1916+0844 & 1175 & 1475 & 3.3  $\pm$ 0.4  &   5    $\pm$   1.5  && 5.7  $\pm$ 0.6  &  3.1  $\pm$  0.7  \\
J1920+1110 & 1175 & 1475 & 3.5  $\pm$ 0.7  &   4.7  $\pm$   1.9  && 4.1  $\pm$ 1    &  3.9  $\pm$  2.1  \\
J1921+1419 & 1175 & 1475 & 1.8  $\pm$ 1.2  &   $\cdots^a$                  && 0.4  $\pm$ 0.4  &   $\cdots^a$                \\
\enddata 
\tablenotetext{a}{Implied values for $\beta$ are unphysically large.}

\end{deluxetable} 


\clearpage
\def\tnm{\tablenotemark}
\begin{deluxetable}{cccccccccc}
\tablecolumns{11} 
\tablewidth{0pc} 
\tablecaption{Estimates of scattering measures and constraints on the properties of clumps.\label{tab:sm}}
\tablehead{ 
\colhead{} & \colhead{} & \colhead{} & \colhead{} & 
\colhead{} & \colhead{} & \colhead{} & \colhead{} &
\colhead{} & \colhead{} \\ 
\colhead{PSR} & \colhead{DM} & \colhead{$D$} & \colhead{$l$} & 
\colhead{$b$} & \colhead{Freq.} & \colhead{SM$_{\rm meas}$} & 
\colhead{SM$_{ne2001}$} & \colhead{$\delta$SM} & \colhead{$F_c(\delta \mbox{DM})^2$} \\
\colhead{} & \colhead{(\dmu)} & \colhead{(kpc)} & \colhead{(deg)} & 
\colhead{(deg)} & \colhead{(MHz)} & \colhead{(\smu)} & \colhead{(\smu)} & 
\colhead{(\smu)} & \colhead{(log)} \\ 
\colhead{(1)} & \colhead{(2)} & \colhead{(3)} & \colhead{(4)} & 
\colhead{(5)} & \colhead{(6)} & \colhead{(7)} & \colhead{(8)} &
\colhead{(9)} & \colhead{(10)} \\ 
}
\startdata 
J1849+0127 & 214.4  & 5.49 &  33.95  &  1.20 &  430  &   0.21  &  0.11  &  0.093  &	--4.04 \\
J1853+0546 & 197.2  & 5.37 &  38.24  &  2.28 & 1175  &   3.58  &  0.033 &  3.54   &     --2.45 \\
           &        &      &         &       & 1475  &   4.79  &        &  4.76   &     --2.32 \\
           &        &      &         &       & 2380  &   7.67  &        &  7.64   &     --2.11 \\
J1855+0422 & 438.6  & 8.39 &  37.22  &  1.20 & 1175  &   4.36  &  0.46  &  3.90   &     --2.60 \\
           &        &      &         &       &       &         &        &         &           \\
J1856+0404 & 345.3  & 6.85 &  37.07  &  0.84 & 1175  &   2.15  &  0.41  &  1.74   &     --2.86 \\
J1857+0526 & 468.3  & 8.92 &  38.40  &  1.24 & 1175  &   2.47  &  0.41  &  2.05   &     --2.90 \\
           &        &      &         &       & 1475  &   2.33  &        &  1.92   &     --2.93 \\
J1858+0215 & 702.0  &10.02 &  35.68  & --0.43 & 1175  &   4.98  &  2.14  &  2.84   &     --2.81 \\
J1901+0413 & 367.0  & 6.81 &  37.77  & --0.20 &  430  &   4.32  &  0.58  &  3.75   &     --2.53 \\
           &        &      &         &       &       &         &        &         &           \\
J1903+0609 & 398.0  & 7.21 &  39.72  &  0.24 & 1175  &   0.73  &  0.59  &  0.14   &     --3.98 \\
J1904+0802 & 438.3  & 8.67 &  41.51  &  0.89 & 1175  &   0.68  &  0.36  &  0.32   &     --3.69 \\
J1905+0616 & 262.7  & 5.76 &  40.05  & --0.15 & 1175  &   3.35  &  0.22  &  3.13   &     --2.53 \\
J1908+0839 & 516.6  & 9.35 &  42.51  &  0.29 & 1175  &   1.07  &  0.57  &  0.51   &     --3.53 \\
J1908+0909 & 464.5  & 8.96 &  42.96  &  0.52 & 1175  &   0.99  &  0.40  &  0.60   &     --3.44 \\
           &        &      &         &       &       &         &        &         &           \\
J1909+0912 & 421.4  & 8.27 &  43.11  &  0.32 & 1175  &   1.11  &  0.44  &  0.67   &     --3.36 \\
J1910+0534 & 484.0  &10.40 &  40.00  & --1.57 & 1175  &   1.92  &  0.19  &  1.73   &     --3.04 \\
J1912+0828 & 359.5  & 7.74 &  42.81  & --0.67 & 1175  &   0.80  &  0.28  &  0.53   &     --3.43 \\
J1913+0832 & 359.5  & 7.93 &  42.98  & --0.86 & 1175  &   1.60  &  0.23  &  1.38   &     --3.03 \\
J1913+1145 & 630.7  &14.56 &  45.83  &  0.63 & 1175  &   1.12  &  0.21  &  0.92   &     --3.47 \\
           &        &      &         &       &       &         &        &         &           \\
J1920+1110 & 181.1  & 5.60 &  46.11  & --1.16 & 1175  &   3.53  &  0.038 &  3.49   &     --2.47 \\
J1852+0031 & 680.0  & 7.91 &  33.46  &  0.11 & 1175  &  51.65  & 44.29  &  7.36   &     --2.30 \\
           &        &      &         &       & 1475  &  61.52  &        & 17.23   &     --1.93 \\
J1902+0556 & 179.7  & 4.70 &  39.41  &  0.36 &  430  &   0.09  &  0.078 &  0.013  &     --4.84 \\
J1909+1102 & 148.4  & 4.17 &  44.74  &  1.17 &  430  &   0.0l7 &  0.013 &  0.0038 &	--5.31 \\
           &        &      &         &       &       &         &        &         &           \\
J1910+0714 & 124.1  & 4.05 &  41.48  & --0.80 &  430  &   0.082 &  0.019 &  0.063  &	--4.07 \\
J1910+1231 & 274.4  & 7.73 &  46.17  &  1.63 &  430  &   0.11  &  0.044 &  0.065  &	--4.34 \\
J1913+1400 & 144.4  & 5.12 &  47.82  &  1.67 &  430  &   0.029 &  0.020 &  0.0089 & 	--5.03 \\
J1915+1606 & 168.8  & 5.91 &  49.91  &  2.22 & 1175  &   0.15  &  0.015 &  0.13   &     --3.91 \\
J1916+1030 & 387.0  & 8.59 &  45.06  & --0.60 & 1175  &   1.74  &  0.21  &  1.54   &     --3.01 \\
           &        &      &         &       &       &         &        &         &           \\
J1926+1434 & 205.0  & 6.44 &  49.80  & --0.84 &  430  &   0.052 &  0.036 &  0.016  &	--4.88 \\
J1853+0505 & 273.6  & 6.49 &  37.64  &  1.97 & 1175  &  19.20  &  0.088 &  19.11  &	--1.80 \\
           &        &      &         &       & 1475  &  22.17  &        &  22.08  &	--1.73 \\
J1915+0856 & 339.0  & 7.96 &  43.56  & --1.11 & 1175  &   1.60  &  0.14  &  1.46   &     --3.00 \\
           &        &      &         &       & 1475  &   1.93  &        &  1.78   &     --2.92 \\
J1935+1745 & 214.6  & 6.93 &  53.63  & --1.21 &  430  &   0.082 &  0.019 &  0.06   &     --4.30 \\
\enddata                                                                                             

\end{deluxetable}                                                                                   

\clearpage
\begin{figure*}
\epsscale{1.8}
\plotone{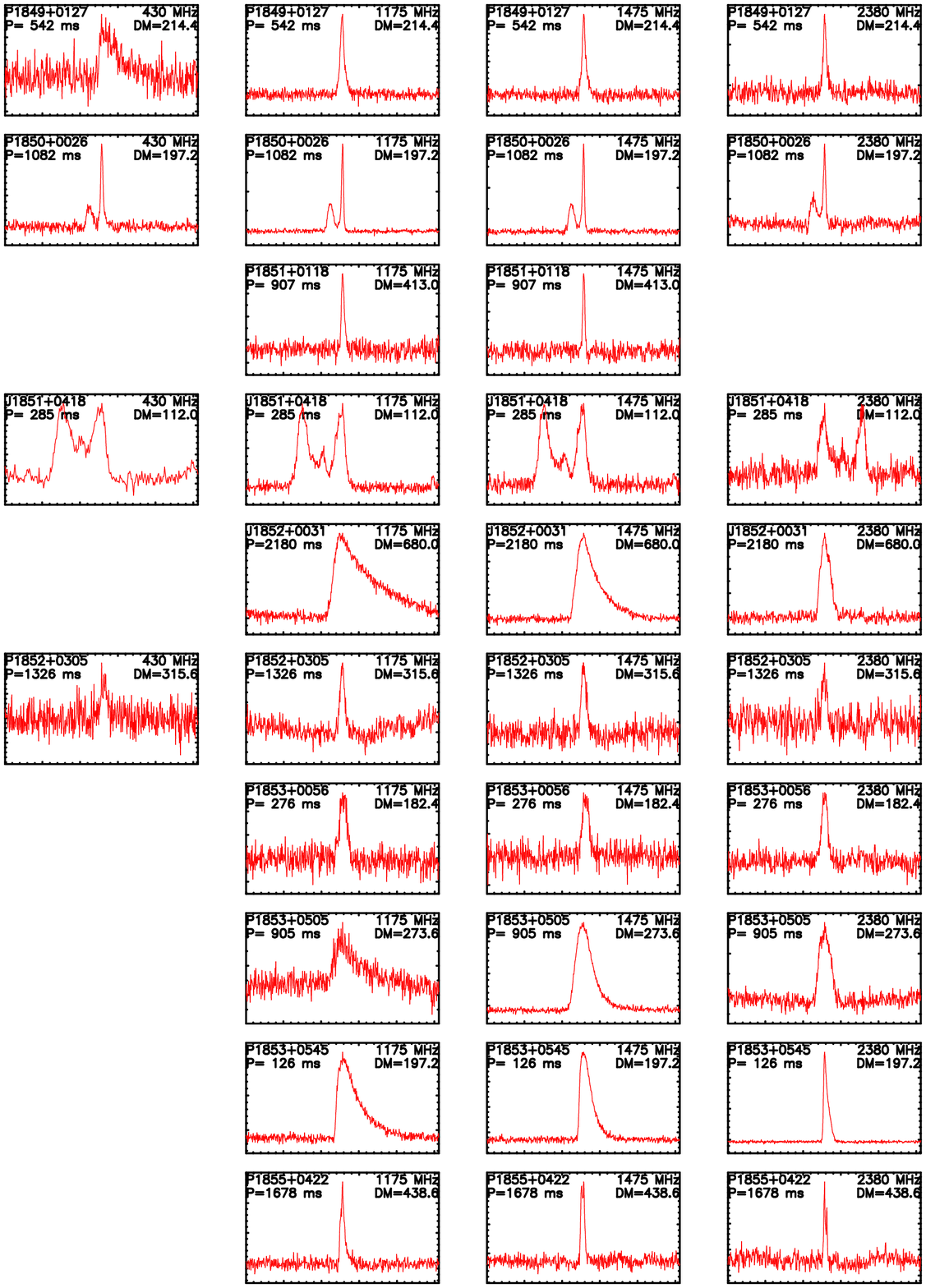}
\end{figure*}

\begin{figure*}
\epsscale{1.8}
\plotone{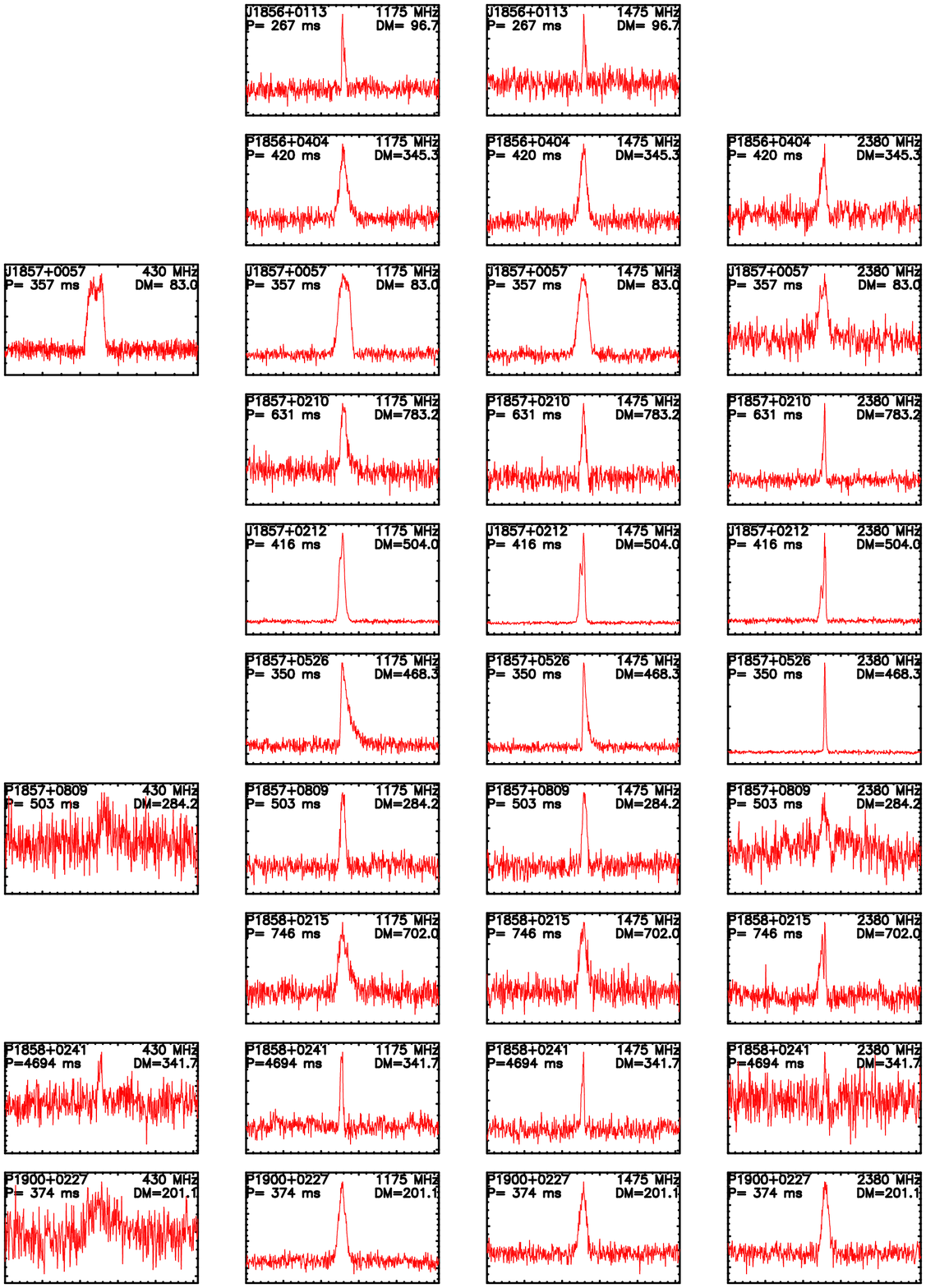}
\end{figure*}

\begin{figure*}
\epsscale{1.8}
\plotone{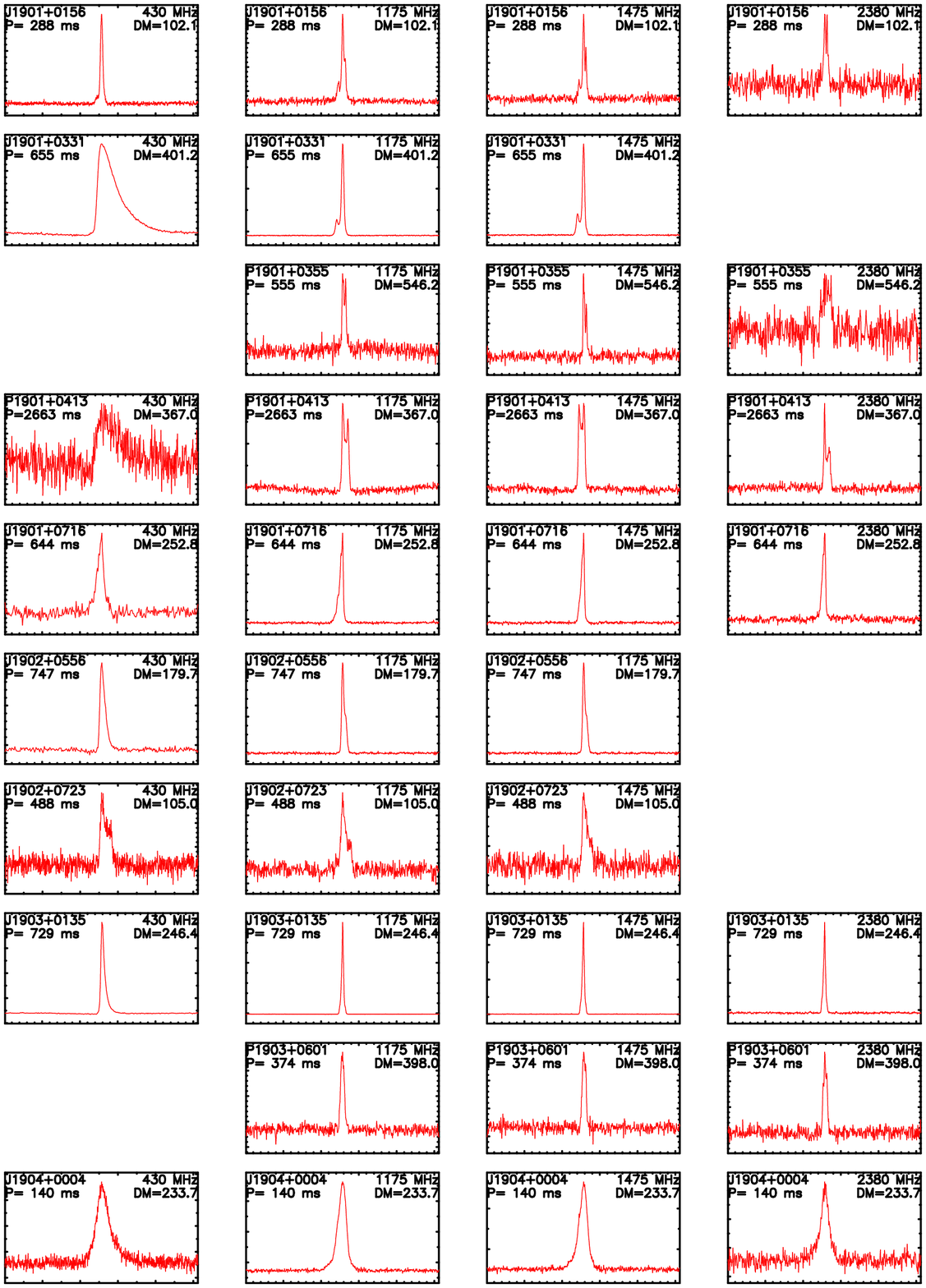}
\end{figure*}

\begin{figure*}
\epsscale{1.8}
\plotone{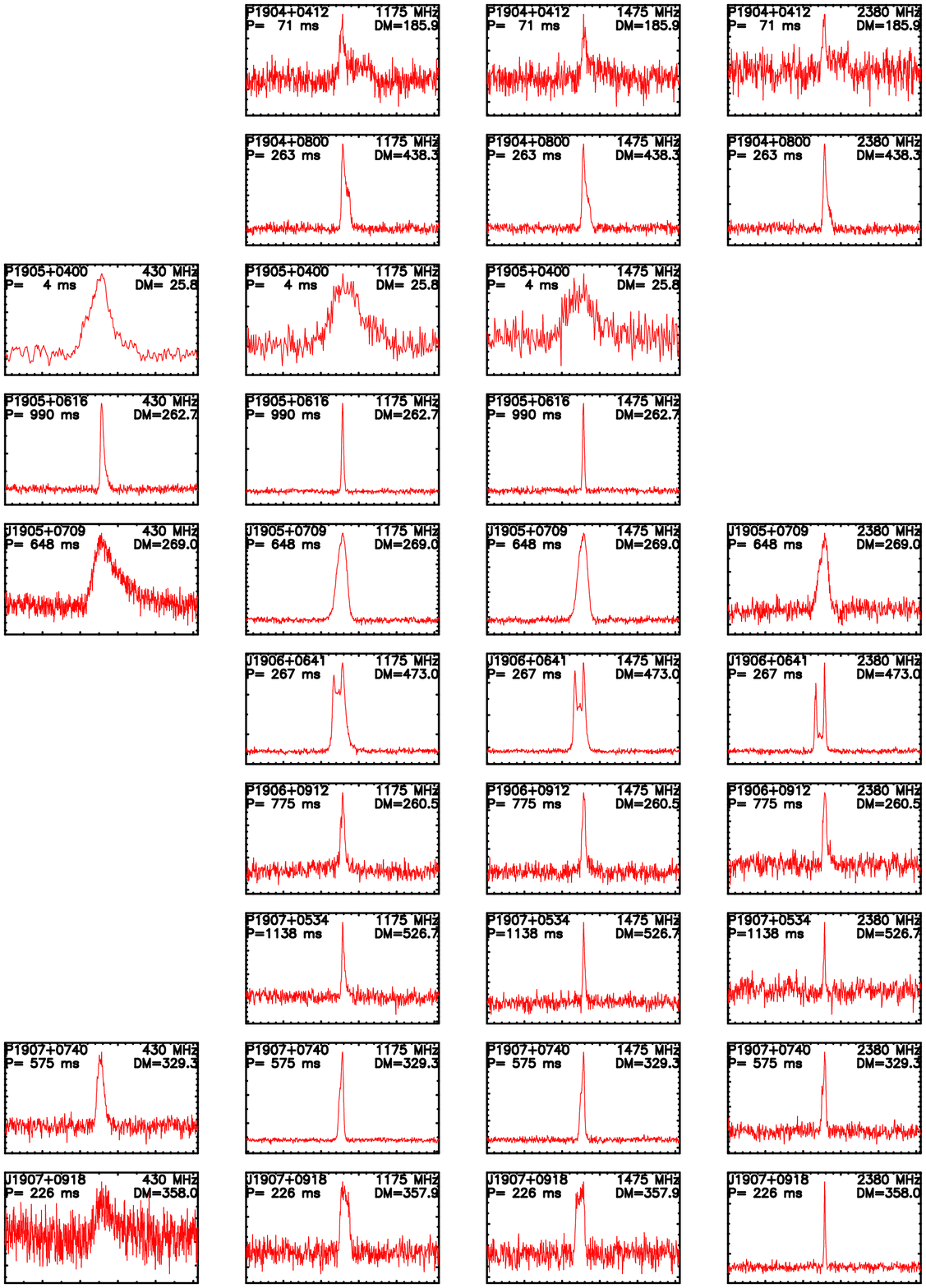}
\end{figure*}

\begin{figure*}
\epsscale{1.8}
\plotone{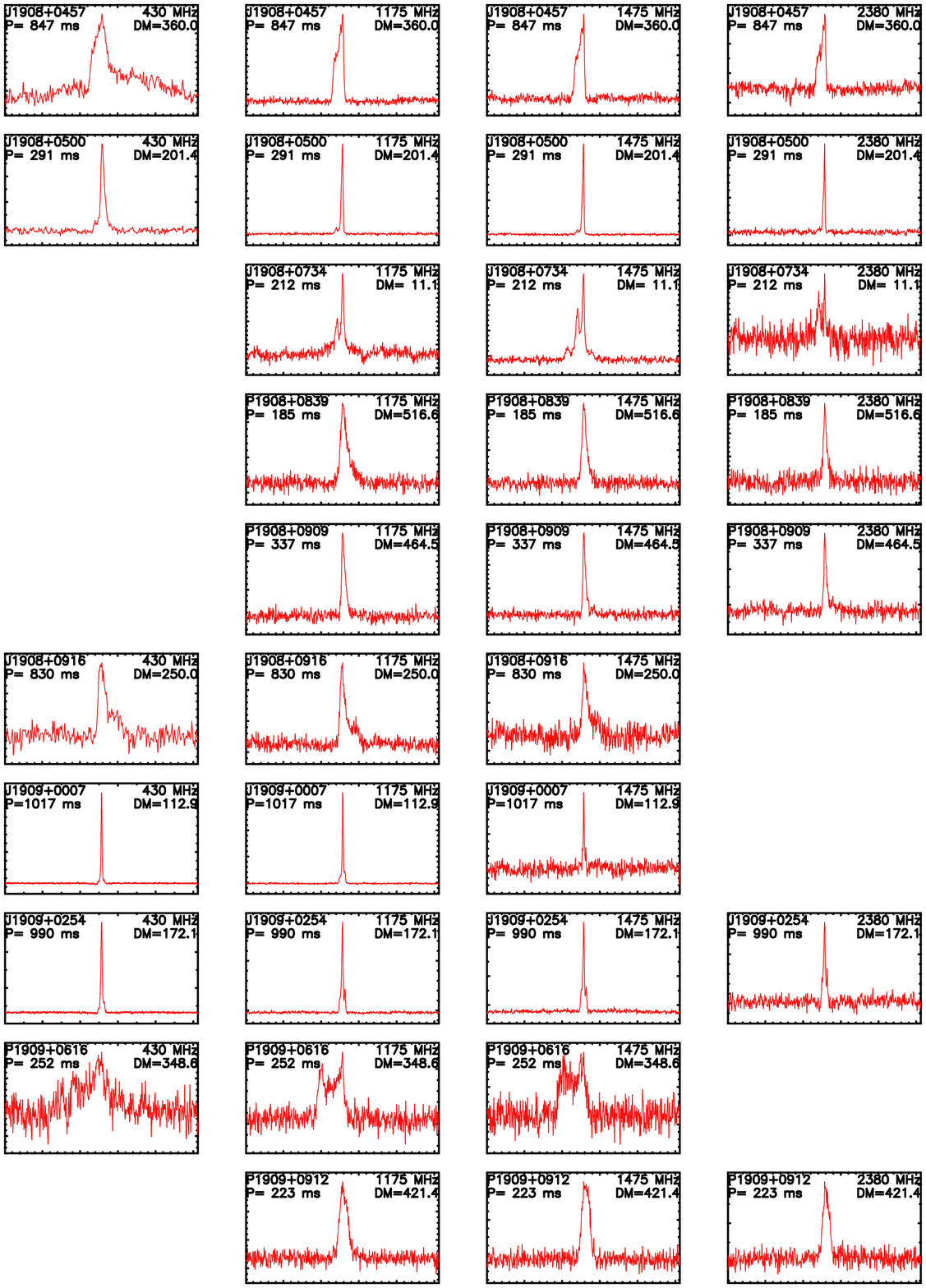}
\end{figure*}

\begin{figure*}
\epsscale{1.8}
\plotone{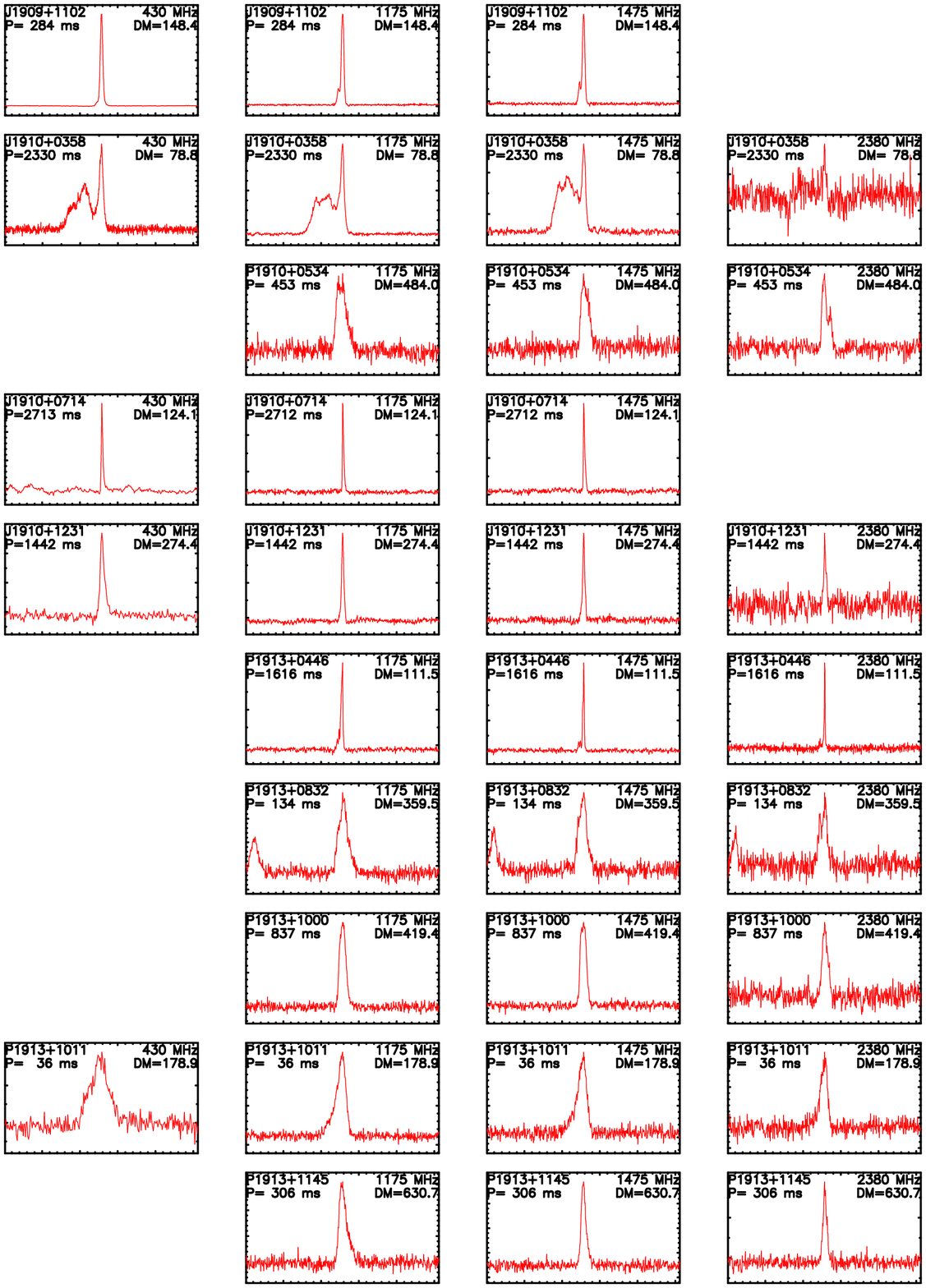}
\end{figure*}

\begin{figure*}
\epsscale{1.8}
\plotone{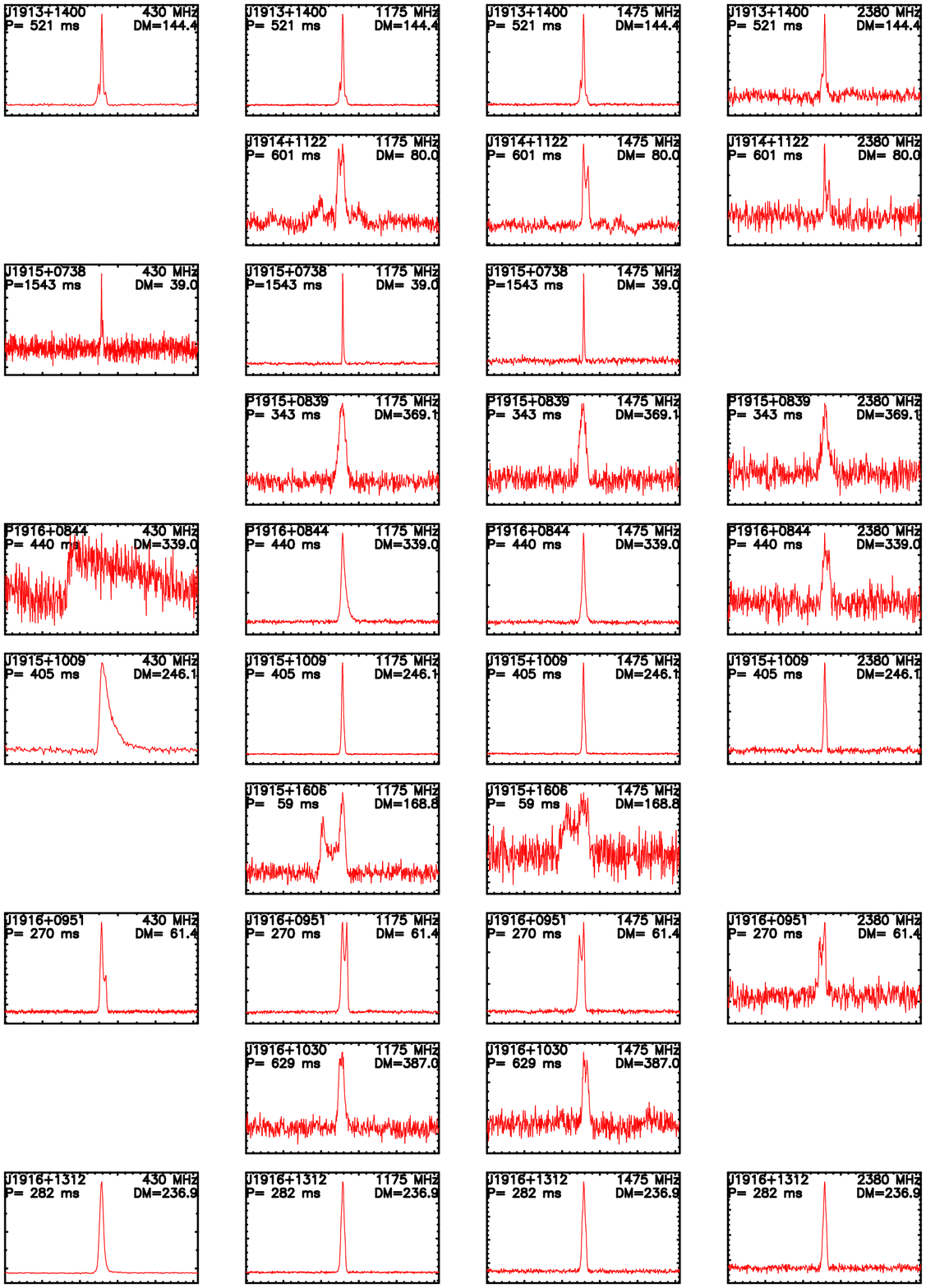}
\end{figure*}

\begin{figure*}
\epsscale{1.8}
\plotone{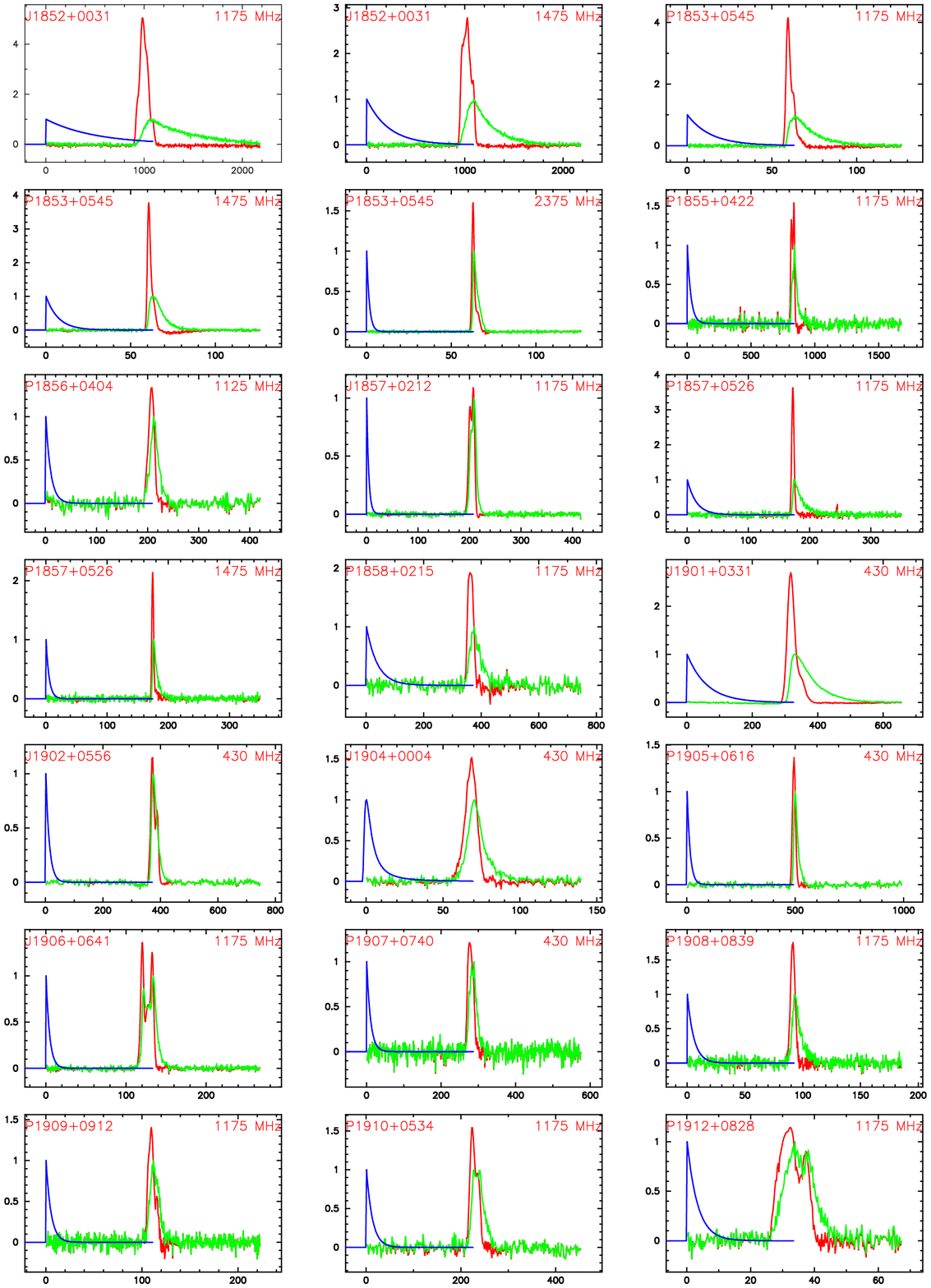}
\end{figure*}

\newpage

\newpage

\clearpage

\end{document}